\documentclass[onecolumn,english,longbibliography,superscriptaddress,breaklinks=true,showkeys,showpacs=false,nofootinbib]{revtex4-2}
\usepackage[T1]{fontenc}
\usepackage[utf8]{inputenc}
\usepackage{color}
\usepackage{babel}
\usepackage{amsmath}
\usepackage{amssymb}
\usepackage{mathtools}
\usepackage{subfigure}
\usepackage{graphicx}
\usepackage{physics} 
\usepackage{dsfont}
\usepackage{xcolor}
\usepackage{ulem}
\usepackage{soul}
\usepackage{cancel}

\usepackage{tikz}
\usepackage{tikz-3dplot}

\usepackage{hyperref}
\hypersetup{
    colorlinks=true,
    linkcolor=blue,
    filecolor=blue,      
    urlcolor=blue,
    }
\makeatletter
\@ifundefined{textcolor}{}
{%
 \definecolor{BLACK}{gray}{0}
 \definecolor{WHITE}{gray}{1}
 \definecolor{RED}{rgb}{1,0,0}
 \definecolor{GREEN}{rgb}{0,1,0}
 \definecolor{BLUE}{rgb}{0,0,1}
 \definecolor{CYAN}{cmyk}{1,0,0,0}
 \definecolor{MAGENTA}{cmyk}{0,1,0,0}
 \definecolor{YELLOW}{cmyk}{0,0,1,0}
}
\pdfoutput=1
\hypersetup{colorlinks=true,citecolor=blue,linkcolor=cyan,urlcolor=blue,filecolor= green, breaklinks=true}
\usepackage{url}
\usepackage{breakurl}
\makeatother
\usepackage[scr=boondox,  
            ]   
           {mathalpha}

\makeatletter
\let\oldHyPsd@CatcodeWarning\HyPsd@CatcodeWarning
\renewcommand{\HyPsd@CatcodeWarning}[1]{
  \ifnum\pdfstrcmp{#1}{math shift}=0    
  \else                                 
    \oldHyPsd@CatcodeWarning{#1}
  \fi
}
\makeatother

\begin{document}
\tdplotsetmaincoords{70}{115} 

\title{Dynamical Casimir effect under the action of gravitational waves}

\author{Gustavo de Oliveira\href{https://orcid.org/0009-0002-2602-971X}{\includegraphics[scale=0.05]{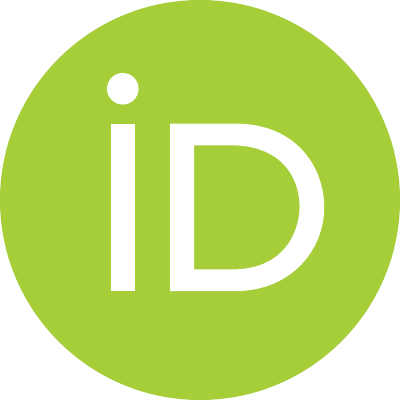}}}
\email{gustav.o.liveira@discente.ufg.br}
\affiliation{QPequi Group, Institute of Physics, Federal University of Goi\'{a}s, 74.690-900, Goi\^{a}nia, Brazil}

\author{Thiago Henrique Moreira\href{https://orcid.org/0000-0001-7093-0287}{\includegraphics[scale=0.05]{orcidid.pdf}}}
\email{thiagoh.moreira6@gmail.com}
\affiliation{QPequi Group, Institute of Physics, Federal University of Goi\'{a}s, 74.690-900, Goi\^{a}nia, Brazil}

\author{Lucas Chibebe C\'{e}leri\href{https://orcid.org/0000-0001-5120-8176}{\includegraphics[scale=0.05]{orcidid.pdf}}}
\email{lucas@qpequi.com}
\affiliation{QPequi Group, Institute of Physics, Federal University of Goi\'{a}s, 74.690-900, Goi\^{a}nia, Brazil}

\begin{abstract}
Several nontrivial phenomena emerge when a quantum field is subjected to dynamical perturbations, with prominent examples including the Hawking and Unruh effects, as well as the dynamical Casimir effect. In this work, we compute the number of particles produced via the dynamical Casimir effect in an ideal cavity, where one of the mirrors is allowed to move under the influence of a gravitational wave. Assuming an oscillatory mirror motion and a plane gravitational wave, we identify the resonance conditions that lead to an exponential increase in the number of created particles through parametric amplification.
\end{abstract}

\maketitle
\section{Introduction}

A variety of remarkable phenomena emerge from the quantum nature of fields. In static settings, vacuum fluctuations give rise to effects such as the Lamb shift~\cite{Lamb1947}, spontaneous emission~\cite{Weisskopf1930}, and the Casimir effect~\cite{Casimir1948}. Under dynamical conditions, even richer behaviour emerges. The Hawking~\cite{hawking1974} and Unruh~\cite{Unruh1976} effects originate from the observer- and time-dependence of the particle concept. In the presence of strong external fields, the vacuum exhibits phenomena such as the Schwinger effect~\cite{Schwinger1951} and vacuum birefringence~\cite{Heisenberg1936}. The topological aspects of the quantum vacuum are exemplified by the Aharonov–Bohm effect~\cite{Aharonov1959}. Finally, time-dependent boundary conditions can induce the creation of particles from vacuum through the dynamical Casimir effect~\cite{Moore1970}, which is the focus of this work.

The dynamical Casimir effect (DCE), which consists of particle creation from the vacuum as a result of a time-dependent boundary condition imposed on the field, has been extensively investigated in a wide range of physical settings. During the past five decades, significant progress has been made, including studies of imperfect mirrors~\cite{Jackel1992,Barton1995,Caloreracos1995,Haro2007}, nontrivial geometries~\cite{Dalvit2006,Mazzitelli2006,Pascoal2008,Pascoal2009,Naylor2012}, nonlinear interactions~\cite{Akopyan2021,Trunin2022,Trunin2023}, entanglement generation and dynamics~\cite{Romualdo2019,Grosso2020}, and thermodynamic irreversibility~\cite{Oliveira2024}. A comprehensive overview of these developments can be found in a recent review~\cite{Dodonov2020}.

Beyond its fundamental significance, the DCE holds substantial physical and practical relevance. It serves as a paradigmatic example of the creation of particles arising from time-dependent boundary conditions~\cite{Moore1970,FullingDavies1976}. Consequently, the DCE functions as a crucial analogue model for quantum field theory in nonstationary backgrounds~\cite{Nation:2011dka,faccio2013analogue}. Within this framework, it provides a controlled laboratory environment to investigate mechanisms analogous to Hawking radiation and cosmological particle production~\cite{Johansson2009}. Furthermore, the effect has achieved successful experimental realisations in superconducting circuits and related platforms~\cite{Wilson2011,Lahteenmaki2013}, systems that enable the simulation of boundary motion at relativistic speeds.

The influence of gravitational fields on the dynamical Casimir effect has been explored, for instance, in Refs.~\cite{Lock2017,Celeri2009}, where the impact of spacetime curvature on particle creation was investigated. In specific, the particle creation induced by the dynamical conditions associated with gravitational waves has also been studied in the context of Bose–Einstein condensates~\cite{Sabin2014}. However, the specific interplay between the mechanical driving of boundaries and the dynamical modulation of spacetime by a gravitational wave remains less explored. That is the focus of the present article.

The central problem addressed in this work is to determine how a classical gravitational wave modifies the resonance conditions and particle creation rates in a cavity that is already subject to mechanical motion. Although standard DCE studies focus on purely mechanical driving, we investigate the specific spectral signature introduced by the gravitational perturbation. Our goal is to isolate the gravitational contribution to the parametric amplification of vacuum fluctuations, distinguishing it from the standard mechanical effect through its unique resonant structure.

While astrophysical gravitational waves are too weak to induce an observable dynamical Casimir effect in current cavities, resonant cavity systems provide a theoretically clean framework for studying how spacetime perturbations couple to quantum vacuum fluctuations. The present work characterizes this coupling at the level of resonance conditions and particle production, and is relevant both for analogue-gravity platforms and for understanding the quantum limits of gravitational-wave–induced field excitations.

This paper is organised as follows. In Sec.~\ref{sec:field}, we derive the field equations for the system interacting with a gravitational wave. The Hamiltonian formulation of the dynamical Casimir effect is reviewed in Sec.~\ref{sec:casimir}. In Sec.~\ref{sec:number}, we present the results for the number of particles created in the field, and Sec.~\ref{sec:conclusions} is devoted to our conclusions.

\section{Scalar field coupled with gravitational waves}
\label{sec:field}

To investigate how gravitational waves influence the creation of particles due to the dynamical Casimir effect, it is crucial to obtain the equations of motion that govern the field interaction with the gravitational wave. Although this is not new, we provide the derivation here to establish the notation and enhance completeness. 

In particular, we examine a real massless scalar field $\Phi(x^{\mu})$ on a curved background described by a $(3+1)$-dimensional manifold $\mathcal{M}$ with a Lorentzian metric $g_{\mu\nu}$. The associated field action is
\begin{equation}
    \label{S}
    S=-\frac{1}{2}\int \dd^4x \sqrt{-g}g^{\mu\nu}\partial_\mu \Phi\partial_\nu \Phi,
\end{equation}
with $g$ being the determinant of the metric. Taking~\eqref{S} to be stationary, one obtains the Klein-Gordon equation
\begin{equation}
    \label{KG}
\frac{1}{\sqrt{-g}}\partial_\nu \left(\sqrt{-g}g^{\mu\nu}\partial_\mu \right)\Phi(x^{\mu})=0.
\end{equation}
As usual, the field conjugate momentum is defined as
$$
\Pi(x^\mu)=-\sqrt{-g}g^{00}\partial_0\Phi(x^\mu).
$$

In the linear approximation, the spacetime metric $g_{\mu\nu}$ can be expanded around a flat metric $\eta_{\mu\nu}$ in terms of a small perturbation $h_{\mu\nu}$, namely:
 \begin{equation}
     g_{\mu\nu} = \eta_{\mu\nu}+h_{\mu\nu},
 \end{equation}
where $|h_{\mu\nu}|\ll 1$. 

We now proceed by considering the field equations outside any source of gravity. In this way, we neglect the backreaction of both the gravitational wave and the field on the spacetime curvature. Under this condition and taking advantage of the gauge invariance of linearised gravity, we choose to work in the transverse-traceless (TT) gauge, so that the metric perturbation obeys $\Bar{h}_{0\nu}=0$, ${\Bar{h}^{\mu}}_{\,\,\,\mu}=0$, and $\partial^\mu\Bar{h}_{\mu\nu}=0$~\cite{Carroll2019}. Under these conditions, Einstein's field equations take the linear form
 \begin{equation}
 \label{eq:EinsteinsFieldEqs}
     \Box\Bar{h}_{\mu\nu}=0,
 \end{equation}
which describe the propagation of gravitational waves $h_{\mu\nu}$ in flat spacetime. The solutions to these equations are of the form
\begin{equation*}
    \Bar{h}_{\mu\nu}(x)=h_{s}\epsilon_{\mu\nu}^se^{ik_\rho x^\rho},
\end{equation*}
where $h_s$ denotes the wave amplitude of polarisation $s$, which is generally referred to as plus ($+$) and cross ($\times$), with reference to the pattern of stretching and compression caused by the wave. $k_\rho$ is the wave vector and $\epsilon_{\mu\nu}^s$ is the polarisation tensor that satisfies the normalisation conditions $\epsilon_{ij}^s(\mathbf{k})\epsilon^{ij}_{s'}(\mathbf{k})=2\delta^s_{s'}$, the transversality $k^i\epsilon_{ij}^{s}(\mathbf{k})=0$, and the traceless $\delta^{ij}\epsilon_{ij}^{s}(\mathbf{k})=0$ conditions.

Putting everything together, the equation of motion for the scalar field in the TT gauge and under the linearised gravity approximation takes the form
\begin{equation}
\label{eq:dynamicaleq}
    \Box\Phi-\Bar{h}_{ij}\partial^i\partial^j\Phi=0,
\end{equation}
where the expansion $\sqrt{-g}=1+h/2+O(h^2)=1+O(h^2)$ was employed. The first term of this equation describes the evolution of the scalar field $\Phi$, while the second term describes its interaction with the gravitational wave. 

In particular, from now on, we will consider, for simplification purposes, that the gravitational wave is travelling in the $z$ direction. In this circumstance, a set of solutions to the wave equation~\eqref{eq:EinsteinsFieldEqs} takes the form of
\begin{subequations} \label{Gravitational-wave-z-direction}
\begin{equation}
    h_{11}(z,t)=-h_{22}(z,t)=h_+\cos[\Omega_g(t-z)],
\end{equation}
\begin{equation}
    h_{12}(z,t)=h_{21}(z,t)=h_\times\cos[\Omega_g(t-z)+\delta_g],
\end{equation}
\end{subequations}
with all other components vanishing. Here, $\Omega_g$ denotes the wave frequency while $\delta_g$ is a phase factor. In the next section, we discuss the dynamical Casimir effect for the scalar field in this context.

\section{The dynamical Casimir effect}
\label{sec:casimir}

In the TT gauge, the coordinate system $(t, x, y, z)$ represents the inertial reference frame of an idealised free-falling observer. This means that test masses, initially at rest, maintain constant coordinate positions to the first order in the metric perturbation $h_{\mu\nu}$. This reflects the fact that gravitational waves induce tidal forces, altering the proper distances between the test masses while leaving their coordinate positions unchanged in this gauge.

\begin{figure}
    \centering
    \includegraphics[width=1\linewidth]{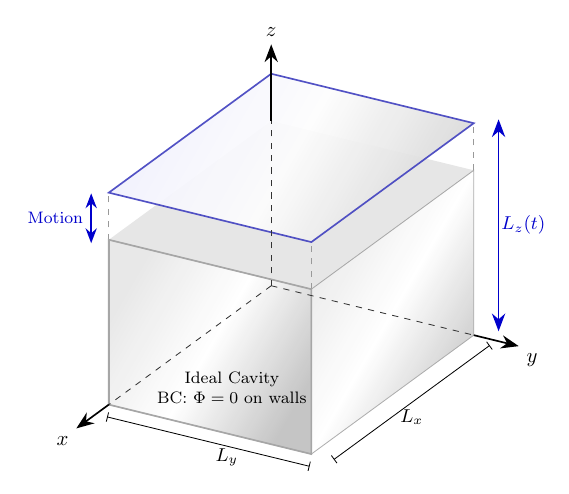}
    \caption{\textbf{The system}. A 3-dimensional cavity that confines the quantum scalar field $\Phi$. We assume that the boundaries of the cavity are formed by perfect mirrors, thus imposing the Dirichlet boundary conditions $\Phi(0,y,z,t)~=~\Phi(L_x,y,z,t)~=~0$, $\Phi(x,0,z,t)~=~\Phi(x,L_y,z,t)~=~0$, and $\Phi(x,y,0,t)~=~\Phi(x,y,L_z(t),t)~=~0$.}
    \label{fig:cavity}
\end{figure}

To study the field dynamics within a cavity under such conditions, we introduce a 3-dimensional ideal cavity in the form of the cuboid $\Sigma(t)=\{\mathbf{x}: 0\leq x \leq L_x, 0\leq y \leq L_y, 0\leq z \leq L_z(t) \}$ that confines the field, as sketched in Fig.~\ref{fig:cavity}. There, the complete system is considered to be in free fall along the $z$-direction, such that the position of the first mirror in this direction is fixed within the coordinate system, while the second mirror is assigned a time-dependent coordinate position $L_z(t)$ .

In practice, implementing a free-falling optical cavity, although difficult, can, in principle, be realised under certain experimental conditions. For example, space-based platforms or suspended optical systems --- such as those used in gravitational wave detectors like LIGO --- can approximate free-fall conditions. 
 
Since we are considering an ideal situation, we impose Dirichlet boundary conditions on all walls of the cavity (mirrors); that is, we demand $\Phi=0$ on the walls at every instant in time. Recall that we have a non-trivial condition on the moving wall, $\Phi(x,y,L_z(t),t)=0$.

Inspecting Eq.~\eqref{eq:dynamicaleq}, it follows that the system must respect the corresponding set of differential equations for $\Phi$ and its conjugate momentum $\Pi$
\begin{subequations}
\label{eq:dyneqn}
    \begin{align}
    \partial_t \Phi(\mathbf{x},t)&=\Pi(\mathbf{x},t),\\
    \partial_t \Pi(\mathbf{x},t)&=\nabla^2 \Phi(\mathbf{x},t)+\Bar{h}_{ij}\partial^i\partial^j\Phi(\mathbf{x},t).
\end{align}
\end{subequations}

The quantum description of the problem is then obtained by promoting the classical functions $\Phi(\mathbf{x},t)$ and $\Pi(\mathbf{x},t)$ to the status of operators $\hat{\Phi}(\mathbf{x},t)$ and $\hat{\Pi}(\mathbf{x},t)$, simultaneously satisfying Eqs.~\eqref{eq:dynamicaleq} and the boundary conditions, as well as the equal-time commutation relations
\begin{subequations}
    \label{eq:commrel}
    \begin{align}
    &\comm{\hat{\Phi}(\mathbf{x},t)}{\hat{\Phi}(\mathbf{y},t)}=\comm{\hat{\Pi}(\mathbf{x},t)}{\hat{\Pi}(\mathbf{y},t)}=0,\\
    &\comm{\hat{\Phi}(\mathbf{x},t)}{\hat{\Pi}(\mathbf{y},t)}=i\delta(\mathbf{x}-\mathbf{y}),
\end{align}
\end{subequations}
with $\mathbf{x},\mathbf{y}\in \Sigma(t)$.

In order to describe the dynamics of the system, we closely follow the formulation given by~\cite{Law1994}, which is based on the introduction of a set of mode functions $\{\varphi_{\mathbf{k}}[\mathbf{x},L_z(t)]\}$ defined through the differential equation
\begin{align}
\label{eq:devarphi}
 \left(\nabla^2 +\Bar{h}_{ij}(\mathbf{x},t)\partial^i\partial^j+\omega_{\mathbf{k}}^2[L_z(t)]\right)\varphi_{\mathbf{k}}[\mathbf{x},L_z(t)]=0, 
\end{align}
subjected to the same dynamical boundary conditions. Here, both the mode function and its eigenfrequency $\omega_{\mathbf{k}}[L_z(t)]$ are defined instantaneously with the mirror position $L_z(t)$. Although $\varphi_{\mathbf{k}}$ and $\omega_{\mathbf{k}}$ depend on time only implicitly through $L_z(t)$, for simplicity of presentation, from now on we will employ the generic time dependence $(t)$ on them.

Up to the second order in the perturbation $h_+$, solutions to Eq.~\eqref{eq:devarphi} can be found in the following form (see Appendix~\ref{AppendixA} for details)
\begin{equation}
    \varphi_{\mathbf{k}}(\mathbf{x},t)=\sqrt{\frac{8}{V(t)}}\sin(k_xx)\sin(k_yy)\sin\left[k_z(t) z\right],
\end{equation}
where $V(t) = L_xL_yL_z(t)$ is the volume of the cavity, while $k_x=n_x\pi/L_x$, $k_y=n_y\pi/L_y$, and $k_z(t)=n_z\pi/L_z(t)$, with $n_x,n_y,n_z\in\mathbb{Z}$, represent the wave numbers in the three spatial directions. The eigenfrequencies are given by
\begin{equation}
    \omega_{\mathbf{k}}^2(t) = k_x^2+k_y^2+k_z^2(t)
    +h_+(k_x^2-k_y^2)\cos(\Omega_g t).
    \label{eq:timefreq}
\end{equation}

Since $\{\varphi_{\mathbf{k}}(\mathbf{x};t)\}$ forms a complete and orthonormal set of mode functions, we can write
\begin{equation}
\label{fieldsexp}
    \hat{\Phi}(\mathbf{x},t)=\sum_{\mathbf{k}} \hat{q}_{\mathbf{k}}(t)\varphi_{\mathbf{k}}(\mathbf{x};t),\qquad 
    \hat{\Pi}(\mathbf{x},t)=\sum_{\mathbf{k}} \hat{p}_{\mathbf{k}}(t)\varphi_{\mathbf{k}}(\mathbf{x};t).   
\end{equation}
Here, the position and momentum field quadratures, $\hat{q}_{\mathbf{k}}(t)$ and $\hat{p}_{\mathbf{k}}(t)$, are defined by
\begin{equation}
\label{eq:qp}    
        \hat{q}_{\mathbf{k}}(t)=\int_{\Sigma(t)}\dd^3 x~\hat{\Phi}(\mathbf{x},t)\varphi_{\mathbf{k}}(\mathbf{x};t),\qquad
        \hat{p}_{\mathbf{k}}(t)=\int_{\Sigma(t)}\dd^3 x~\hat{\Pi}(\mathbf{x},t)\varphi_{\mathbf{k}}(\mathbf{x};t),
\end{equation}
and, as a result of Eqs.~\eqref{eq:commrel}, they satisfy the following commutation relations
\begin{equation}
    \comm{\hat{q}_{\mathbf{k}}(t)}{\hat{p}_{\mathbf{j}}(t)}=\comm{\hat{p}_{\mathbf{k}}(t)}{\hat{p}_{\mathbf{j}}(t)}=0,\qquad
    \comm{\hat{q}_{\mathbf{k}}(t)}{\hat{p}_{\mathbf{j}}(t)}=i\delta_{\mathbf{k},\mathbf{j}}.
\end{equation}
By differentiating Eqs.~\eqref{eq:qp} with respect to time (and using Eqs.~\eqref{eq:dyneqn},~\eqref{eq:devarphi}, and the orthonormalization condition for the mode functions), one can obtain the following expressions
\begin{equation}
\label{eq:dqdp}
        \frac{\dd \hat{q}_{\mathbf{k}}}{\dd t}=\hat{p}_{\mathbf{k}}+\sum_{\mathbf{j}}G_{\mathbf{k},\mathbf{j}}(t)\hat{q}_{\mathbf{j}},\qquad
        \frac{\dd \hat{p}_{\mathbf{k}}}{\dd t}=-\omega_{\mathbf{k}}^2(t)\hat{q}_{\mathbf{k}}-\sum_{\mathbf{j}}G_{\mathbf{j},\mathbf{k}}(t)\hat{p}_{\mathbf{j}},
\end{equation}
where the time-dependent and anti-symmetric coupling coefficient takes the form
\begin{align}
    G_{\mathbf{k},\mathbf{j}}(t)&:=-\int_{\Sigma(t)}\dd^3x~\varphi_{\mathbf{k}}\dot{\varphi}_{\mathbf{j}}=g_{\mathbf{k},\mathbf{j}}\frac{\dot{L}_z(t)}{L_z(t)},
\end{align}
with
\begin{equation}
g_{\mathbf{k},\mathbf{j}}=
\begin{cases}
    (-1)^{j_z-k_z}\frac{2k_z j_z}{j_z^2-k_z^2}\frac{\dot{L}(t)}{L(t)}\delta_{k_xj_x}\delta_{k_y,j_y},&\, k_z=j_z,\\
    0,&\, k_z\neq j_z.\\
\end{cases}
\end{equation}

To define the Fock space for the system, we further introduce the instantaneous creation and annihilation operators
\begin{subequations}
\label{eq:aad}
    \begin{align}
        \hat{a}_{\mathbf{k}}^{\dagger}(t)&e^{i\Theta_{\mathbf{k}}(t)}=\frac{1}{\sqrt{2\omega_{\mathbf{k}}(t)}}\left[\omega_{\mathbf{k}}(t)\hat{q}_{\mathbf{k}}-i\hat{p}_{\mathbf{k}}(t)\right],\\
        \hat{a}_{\mathbf{k}}(t)&e^{-i\Theta_{\mathbf{k}}(t)}=\frac{1}{\sqrt{2\omega_{\mathbf{k}}(t)}}\left[\omega_{\mathbf{k}}(t)\hat{q}_{\mathbf{k}}+i\hat{p}_{\mathbf{k}}(t)\right],
    \end{align}
\end{subequations}
where we have introduced the time-dependent integrated frequency $\Theta_{\mathbf{k}}(t)=\int_0^t dt^{\prime}\omega_{\mathbf{k}}(t')$ for later convenience. The introduction of this term at this stage has the effect of obtaining the effective Hamiltonian already in the interaction picture. $\hat{a}_{\mathbf{k}}(t)$ and $\hat{a}_{\mathbf{k}}^{\dagger}(t)$ satisfy the standard commutation relations
\begin{equation}
        \Big[\hat{a}_{\mathbf{k}}(t),\hat{a}_{\mathbf{j}}(t)\Big]=     \comm{\hat{a}_{\mathbf{k}}^{\dagger}(t)}{\hat{a}_{\mathbf{j}}^{\dagger}(t)}=0,\qquad
        \comm{\hat{a}_{\mathbf{k}}(t)}{\hat{a}_{\mathbf{j}}^{\dagger}(t)} = \delta_{\mathbf{k}\mathbf{j}}.
\end{equation}

Here, the name instantaneous refers to the physical interpretation that if we freeze the system at some instant $t_0$, the operators $\hat{a}_k(t_0)$ and $\hat{a}_k^{\dagger}(t_0)$ must describe the particle notion for the field as if the cavity mirror had stopped at position $L(t_0)$.

Taking the time derivative of Eqs.~\eqref{eq:aad}, one obtains the following
\begin{subequations}
\label{eq:odeaad}
    \begin{align}
    \frac{\dd \hat{a}_{\mathbf{k}}}{\dd t}&=\sum_{\mathbf{j}}\left[\mu_{[\mathbf{k},\mathbf{j}]}\hat{a}_{\mathbf{j}}^{\dagger}e^{i[\Theta_{\mathbf{k}}+\Theta_{\mathbf{j}}]}+\mu_{(\mathbf{k},\mathbf{j})}\hat{a}_{\mathbf{j}}e^{-i[\Theta_{\mathbf{k}}-\Theta_{\mathbf{j}}]}\right],\\
    \frac{\dd \hat{a}_{\mathbf{k}}^{\dagger}}{\dd t}&=\sum_{\mathbf{j}}\left[\mu_{[\mathbf{k},\mathbf{j}]}\hat{a}_{\mathbf{j}}e^{-i[\Theta_{\mathbf{k}}+\Theta_{\mathbf{j}}]}+\mu_{(\mathbf{k},\mathbf{j})}\hat{a}_{\mathbf{j}}^{\dagger}e^{i[\Theta_{\mathbf{k}}-\Theta_{\mathbf{j}}]}\right],
\end{align}
\end{subequations}
where $\mu_{(\mathbf{k},\mathbf{j})}=\frac{1}{2}\left(\mu_{\mathbf{k},\mathbf{j}}+\mu_{\mathbf{j},\mathbf{k}}\right)$, $\mu_{[\mathbf{k},\mathbf{j}]}=\frac{1}{2}\left(\mu_{\mathbf{k},\mathbf{j}}-\mu_{\mathbf{j},\mathbf{k}}\right)$, and
\begin{equation}
\label{eq:mukj}
   \mu_{\mathbf{k},\mathbf{j}}(t)= \begin{cases}
\frac{1}{2}\frac{\dot{\omega}_{\mathbf{k}}(t)}{\omega_{\mathbf{k}(t)}},&\qquad \mathbf{k}=\mathbf{j},\\
g_{\mathbf{k},\mathbf{j}}\frac{\dot{L}_z(t)}{L_z(t)}\sqrt{\frac{\omega_{\mathbf{k}}(t)}{\omega_{\mathbf{j}}(t)}},&\qquad \mathbf{k}\neq\mathbf{j}.
    \end{cases}
\end{equation}

By interpreting Eqs.~\eqref{eq:odeaad} as the Heisenberg equations of motion $\dd\hat{O}/\dd t = i\comm{\hat{H}_{\text{eff}}}{\hat{O}}$, for $\hat{a}_{\mathbf{k}}(t)$ and $\hat{a}_{\mathbf{k}}^{\dagger}(t)$, one can construct an effective Hamiltonian.
\begin{eqnarray}
    \nonumber\hat{H}_{\text{eff}}(t) &=& \frac{i}{2}\sum_{\mathbf{k}} \mu_{[\mathbf{k},\mathbf{k}]}(t)\left(\hat{a}_{\mathbf{k}}^{\dagger 2}e^{2i\Theta_{\mathbf{k}}(t)}-\hat{a}_{\mathbf{k}}^2e^{-2i\Theta_{\mathbf{k}}(t)}\right) \nonumber \\&+&\frac{i}{2}\sum_{\mathbf{j}\neq\mathbf{k}} \Bigg[\mu_{[\mathbf{k},\mathbf{j}]}(t)\left(\hat{a}_{\mathbf{j}}^{\dagger}\hat{a}_{\mathbf{k}}^{\dagger}e^{i[\Theta_{\mathbf{k}}(t)+\Theta_{\mathbf{j}}(t)]}-\hat{a}_{\mathbf{j}}\hat{a}_{\mathbf{k}}e^{-i[\Theta_{\mathbf{k}}(t)+\Theta_{\mathbf{j}}(t)]}\right)\nonumber\\
   &+&\mu_{(\mathbf{k},\mathbf{j})}(t)\left(\hat{a}_{\mathbf{j}}^{\dagger}\hat{a}_{\mathbf{k}}e^{-i[\Theta_{\mathbf{k}}(t)-\Theta_{\mathbf{j}}(t)]}-\hat{a}_{\mathbf{k}}^{\dagger}\hat{a}_{\mathbf{j}}e^{i[\Theta_{\mathbf{k}}(t)-\Theta_{\mathbf{j}}(t)]}\right)\Bigg],
    \label{eq:effham}
\end{eqnarray}
where $\hat{a}_{\mathbf{k}}(t=0)=\hat{a}_{\mathbf{k}}$. Since the operators in \eqref{eq:aad} incorporate the integrated frequency $\Theta_{\mathbf{k}}(t)$, the resulting effective Hamiltonian \eqref{eq:effham} is inherently expressed in the interaction picture. In this representation, the standard free-energy terms proportional to $\hat{a}_{\mathbf{k}}^{\dagger}\hat{a}_{\mathbf{k}}$ are naturally absorbed into the definition of the operators. See Appendix~\ref{AppendixB} for details of the calculation.

From Eq.~\eqref{eq:effham}, one can see the existence of two different contributions to the dynamical Casimir effect. First, there is a scattering process governed by terms that contain operators $\hat{a}_{\mathbf{k}}^{\dagger}\hat{a}_{\mathbf{j}}$, where particles are scattered from one mode to another without changing the total number of particles in the field. Secondly, we have a pair-creating process characterised by the terms $\hat{a}_{\mathbf{k}}^{\dagger}\hat{a}_{\mathbf{j}}^{\dagger}$ and $\hat{a}_{\mathbf{k}}^{\dagger 2}$, where particles are created even from the vacuum in pairs of different and equal frequencies.  

In the next section, we will discuss how to calculate the number of particles created in a certain unperturbed mode of the field using the aforementioned effective Hamiltonian.

\section{Number of created particles}
\label{sec:number}

The defining feature of DCE is captured by the process of particle-creation as a result of time-dependent changes imposed on the field. We consider that the perturbation acts on the cavity during the time interval $t\in[0,T]$. After this, the cavity returns to its initial position. Using the effective Hamiltonian~\eqref{eq:effham}, the number of particles created in the $\mathbf{k}$-th field mode is given by
\begin{equation}
\nonumber N_{\mathbf{k}}(T)= \bra{0;\text{in}}\hat{a}_{\mathbf{k}}^{\dagger}(T)\hat{a}_{\mathbf{k}}(T)\ket{0;\text{in}}
\label{eq:NkT}=\bra{0;\text{in}}\hat{U}^{\dagger}(0,T)\hat{a}_{\mathbf{k}}^{\dagger}(0)\hat{a}_{\mathbf{k}}(0)\hat{U}(0,T)\ket{0;\text{in}},
\end{equation}
where $\ket{0,\text{in}}$ is the vacuum state prepared initially; whereas 
\begin{equation}
\hat{U}(0,T)=\mathcal{T}\exp{-i\int_{0}^{T}\dd t\hat{H}_{\text{eff}}(t)}
\end{equation}
is the time evolution operator, with $\mathcal{T}$ being the time order operator.

Expanding the final operators $\hat{a}_{\mathbf{k}}(T)$ and $\hat{a}_{\mathbf{k}}^{\dagger}(T)$ in terms of a linear combination of the initial operators $\hat{a}_{\mathbf{k}}(0)$ and $\hat{a}_{\mathbf{k}}^{\dagger}(0)$, one obtains the following
\begin{subequations}
\label{eq:bogol}
\begin{align}
    \hat{a}_{\mathbf{k}}(T)&=\sum_{\mathbf{j}}\left[\alpha_{\mathbf{j}\mathbf{k}}(T)\hat{a}_{\mathbf{j}}(0)+\beta_{\mathbf{j}\mathbf{k}}^*(T)\hat{a}_{\mathbf{j}}^{\dagger}(0)\right],\\
    \hat{a}_{\mathbf{k}}^{\dagger}(T)&=\sum_{\mathbf{j}}\left[\beta_{\mathbf{j}\mathbf{k}}(T)\hat{a}_{\mathbf{j}}(0)+\alpha_{\mathbf{j}\mathbf{k}}^{*}(T)\hat{a}_{\mathbf{j}}^{\dagger}(0)\right],
\end{align}
\end{subequations}
where $\alpha_{\mathbf{j}\mathbf{k}}(T)$ and $\beta_{\mathbf{j}\mathbf{k}}(T)$ are the Bogoliubov coefficients.

By substituting the expansions~\eqref{eq:bogol} into Eq.~\eqref{eq:NkT} and considering the defining property of the vacuum state in terms of the initial operators, i.e. $\hat{a}_{\mathbf{k}}(0)\ket{0,\text{in}}=0$, it can be shown that
\begin{equation}
    N_{\mathbf{k}}(T)=\sum_{\mathbf{j}}|\beta_{\mathbf{k}\mathbf{j}}(T)|^2,
\end{equation}
meaning that whenever the mirror motion leads to a non-vanishing $\beta_{\mathbf{k}\mathbf{j}}(T)$, particles will be created in the field. For explicit expressions for the Bogoliubov coefficients, we encourage the reader to check Appendix~\ref{AppendixBogoliubov}, where they are obtained in terms of an expansion of the time-dependent Hamiltonian coefficients.

By assuming the cavity field to be weakly perturbed periodically by the motion of one of its mirrors, one can select specific field modes to be excited by imposing resonance conditions. For this purpose, we consider the second mirror in the $z$-direction to perform weak oscillations in the form
\begin{equation}
    L_z(t)=L_z[1+\epsilon \sin (\Omega_c t)],
\end{equation}
where $\epsilon \ll 1$ is a small amplitude and $\Omega_c$ is the oscillation frequency of the cavity.

In the case of one-dimensional cavities, the unperturbed field frequencies are all proportional to $\pi/L$, making them equidistant. This has the effect that, whenever one mode is resonantly excited, there will generally always be another mode that will strongly couple with the first. Because of the equidistant property, one can guarantee a chain of modes coupled with each other. This is best exemplified in the case (without gravitational waves) where the cavity is set to oscillate with double the $k$-th unperturbed field frequency $\omega_k$. In such a situation, it can be demonstrated that, based on the parametric creation of particles in the mode $k$, this mode will strongly couple to the $(k+2)$-th mode, scattering some initial particles into it. This thread continues with the mode $k+2$ coupling to the $(k+4)$-th mode, and so forth, thus effectively creating a chain of coupled modes in a cascade of particle creation.

However, for a three-dimensional cavity, as in our case, the spectrum of eigenfrequencies is non-equidistant, meaning that one rarely finds modes whose frequencies are multiples of one another. The consequence is that, in general, we do not need to worry about a sequence of strongly-coupled modes, as in the one-dimensional case. For this matter, we suppose a cavity configuration in which at most only a pair of modes can couple with each other. Using this configuration along with the rotating-wave approximation, one can simplify the Hamiltonian considerably and obtain analytical expressions for the number of particles found at a given field mode. 

Calculating expressions for the number of particles created from the vacuum (see Appendix~\ref{AppendixCalc}), one finds the existence of 4 different resonance conditions of gravitational origin that substantially enhance particle production (Eqs.~\eqref{eq:NkT1}--\eqref{eq:NkT4}), alongside the standard DCE resonance of purely mechanical origin (Eq.~\eqref{eq:NkT}). There, the number of particles $N_{\mathbf{k}}$ that are created in the $\mathbf{k}$-th field mode can, under all resonance conditions, be expressed in the form
\begin{equation}
    N_{\mathbf{k}}(T)=\sinh^2\left(\chi_{\mathbf{k}}T\right),
\end{equation}
where $\chi_{\mathbf{k}}$ is the rate of (exponential) growth of particles due to parametric amplification of vacuum fluctuations induced, in this case, by mechanical and gravitational influences~\cite{Dodonov:2010zza,Nation2012}. For each different resonance condition and whenever $k_x=j_z$ and $k_y=j_y$, one can write $\chi_{\mathbf{k}}$ as follows
\begin{subequations}
\begin{align}
\omega_{\mathbf{k}}=\frac{\Omega_g}{2}:&\qquad  \chi_{\mathbf{k}}=\left|\frac{h_+}{8}\frac{\Omega_g}{\omega_{\mathbf{k},0}^2}(k_x^2-k_y^2)\right|,\label{eq:NkT1}\\
  \omega_{\mathbf{k}}=\frac{|\Omega_c\pm\Omega_g|}{2}:&\qquad \chi_{\mathbf{k}}=\left|\frac{\epsilon h_+}{8}\frac{(k_x^2-k_y^2)k_z^2}{\omega_{\mathbf{k},0}^3} \frac{\Omega_c^2+\Omega_g^2}{\Omega_c\Omega_g}\right|,\label{eq:NkT2}\\
\omega_{\mathbf{k}}+\omega_{\mathbf{j}}=\Omega_g\pm\Omega_c:&\qquad \chi_{\mathbf{k}}=\left|\frac{h_+\epsilon}{16\omega_{\mathbf{k},0}\omega_{\mathbf{j},0}}\frac{\Omega_c}{\Omega_g}\frac{(k_x^2-k_y^2)k_z j_z}{\sqrt{\omega_{\mathbf{k},0}\omega_{\mathbf{j},0}}}\left(1+\frac{\Omega_g}{2}\frac{\Omega_g\pm\Omega_c}{\omega_{\mathbf{k},0}\omega_{\mathbf{j},0}}\right)\right|,\label{eq:NkT3}\\
\omega_{\mathbf{k}}+\omega_{\mathbf{j}}=\Omega_c-\Omega_g:&\qquad \chi_{\mathbf{k}}=\left|\frac{h_+\epsilon}{16\omega_{\mathbf{k},0}\omega_{\mathbf{j},0}}\frac{\Omega_c}{\Omega_g}\frac{(k_x^2-k_y^2)k_z j_z}{\sqrt{\omega_{\mathbf{k},0}\omega_{\mathbf{j},0}}}\left(1-\frac{\Omega_g}{2}\frac{\Omega_c-\Omega_g}{\omega_{\mathbf{k},0}\omega_{\mathbf{j},0}}\right)\right|,\label{eq:NkT4}\\
\label{eq:NkT}\omega_{\mathbf{k}}=\frac{\Omega_c}{2}:&\qquad  \chi_{\mathbf{k}}=\left|\frac{k_z^2 \Omega_c \epsilon}{4 \omega_{\mathbf{k},0}^2}\right|,
\end{align}
\end{subequations}
where $\omega_{\mathbf{k},0}^2=k_x^2+k_y^2+k_z^2(0)$. 

An important aspect of the above results is that Eq.~\eqref{eq:NkT1} matches the exact expression obtained in Ref.~\cite{Barbado:2018qod}. In that context, the authors calculated the number of particles created from a scalar field confined in a static cavity (more specifically, a phonon field of a trapped Bose-Einstein condensate) while it was being perturbed by gravitational waves. Since Eq.~\eqref{eq:NkT1} corresponds to the only resonant condition where the cavity is static, it can be regarded as a consistency check for the validity of our calculations.

The dependence on the gravitational wave amplitude $h_+$, which ranges from $10^{-20}$ to $10^{-21}$~\cite{LIGOScientific:2016aoc, Maggiore:2007ulw} in typical gravitational interferometer experiments, makes the contribution to the number of particles created in Eq.~\eqref{eq:NkT1} extremely small. For the remaining cases in Eqs.~\eqref{eq:NkT}, the contribution is even smaller since the corresponding expressions depend on both $h_+$ and the small mirror oscillating amplitude $\epsilon$, defined as the ratio of the mirror's oscillation amplitude to the total cavity size. Physically reasonable values for $\epsilon$, as found in the literature, typically range from $10^{-6}$ to $10^{-11}$~\cite{Dodonov:2010zza,Johansson:2009zz,Wilson2011}. Regarding realistic frequency scales, mechanical driving frequencies in circuit Quantum Electrodynamics (cQED) are usually in the microwave regime (GHz), while astrophysical gravitational waves detected by ground-based interferometers have frequencies in the audio band (kHz). This creates a clear scale separation where $\Omega_g \ll \Omega_c$, ensuring the validity of the long-wavelength approximation $L_z \Omega_g \ll 1$.

Although the mechanical signal is orders of magnitude larger, it is confined to the frequency $\Omega_c/2$. In contrast, the gravitational signal appears as distinct neighbouring peaks shifted by $\pm \Omega_g/2$. To distinguish these signals, the cavity's resonance must be extremely sharp, requiring a linewidth narrower than the frequency gap $|\Omega_g/2|$. For a GHz cavity and a kHz gravitational wave, this implies a quality factor $Q \gtrsim 10^7$. Under these conditions, the mechanical background can be effectively suppressed at the sideband frequencies $\omega_{\mathbf{k}} = (\Omega_c \pm \Omega_g)/2$, leaving the gravitational signal as the theoretically dominant source of resonant particle production. Although this offers a mechanism for spectral separation from the mechanical input, it does not account for other noise sources like thermal fluctuations, which could still populate these sidebands.

Furthermore, even if spectral resolution is achieved, the absolute intensity of the gravitational signal remains fundamentally limited by the small value of $h_+$. Since the particle production rate scales with the square of the amplitude, the gravitational contribution is suppressed by a factor of roughly $h_+^2 \sim 10^{-42}$ compared to the standard mechanical effect. Thus, in principle, the primary barrier to detection is not the interference from mechanical input, but the extreme sensitivity required to register such a faint signal.

In this context, we clarify that our primary objective in this work is to theoretically isolate and characterise the specific contribution of the gravitational wave to the particle creation process. We acknowledge that under realistic experimental conditions (where $\Omega_g \ll \Omega_c$), the standard dynamical Casimir effect driven by the mechanical motion of the mirrors would overwhelmingly dominate the total particle count. However, the gravitational interaction induces unique resonant signatures (at sideband frequencies), that are distinct from the purely mechanical signal. For this reason, we have neglected higher-order terms in $\epsilon$ that describe standard mechanical particle creation. This approximation allows us to study the perturbative gravitational effect in a conceptually clean manner, treating it as a distinct signal to be resolved rather than merely a subleading correction to the mechanical background.

\begin{figure*} 
\centering
    \subfigure[$2\omega_{\mathbf{k}}=\Omega_c+\Omega_g$]{
        \includegraphics[width=0.475\linewidth]{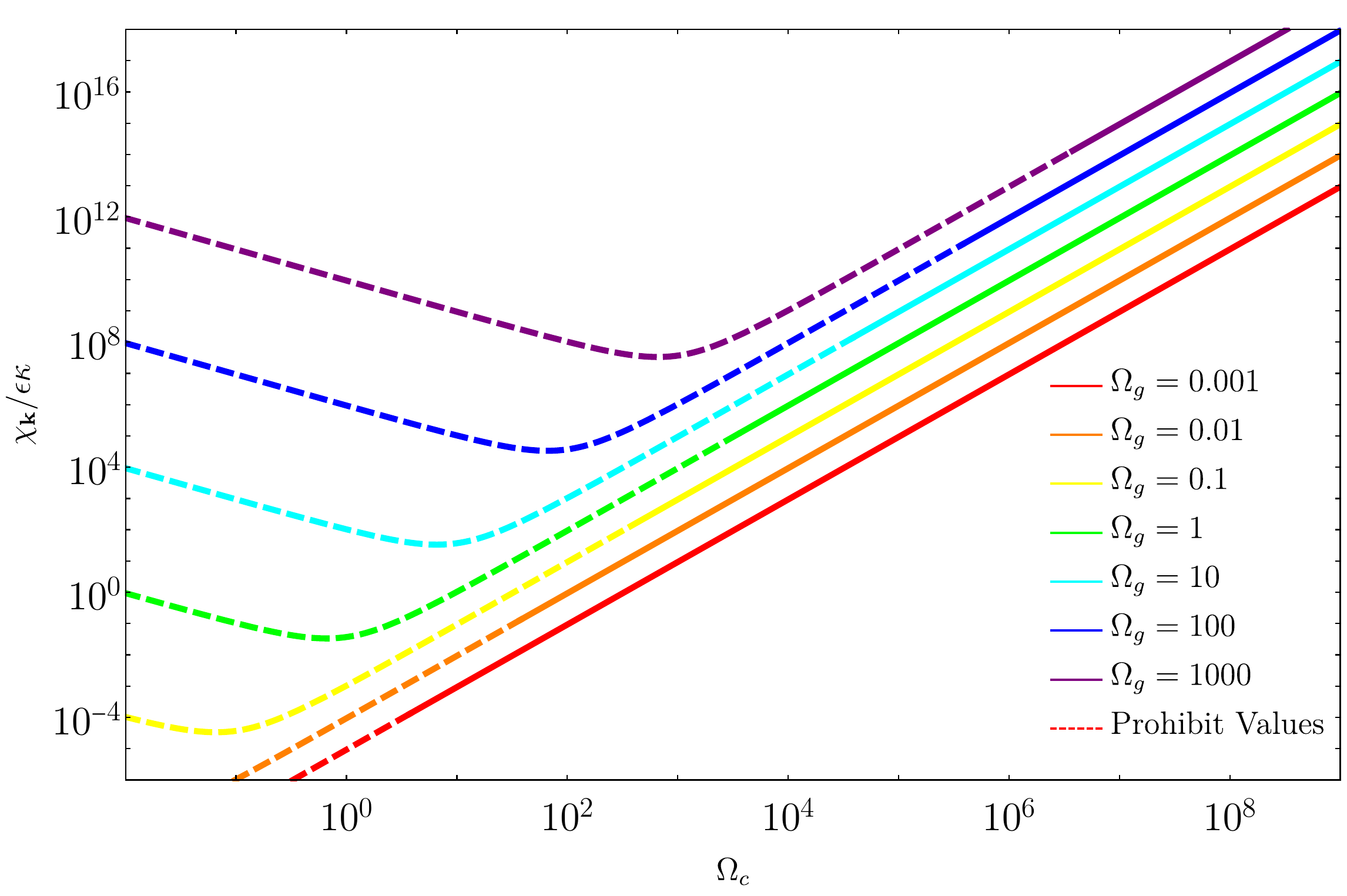} 
    } \hfill 
    \subfigure[$2\omega_{\mathbf{k}}=\Omega_c-\Omega_g$]{
        \includegraphics[width=0.475\linewidth]{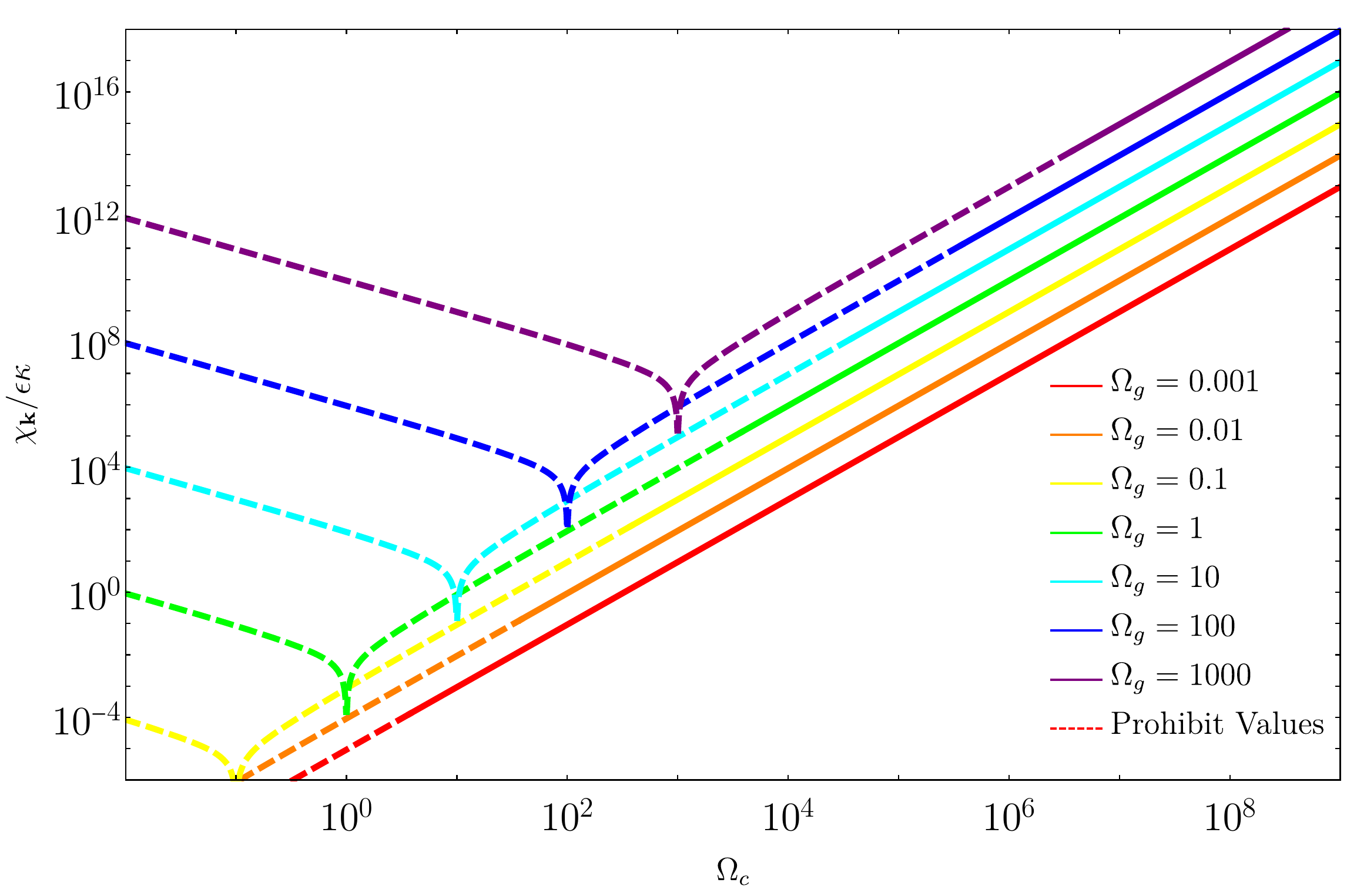}
    } \\ 

    \subfigure[$\omega_{\mathbf{k}}+\omega_{\mathbf{j}}=\Omega_c+\Omega_g$]{
        \includegraphics[width=0.475\linewidth]{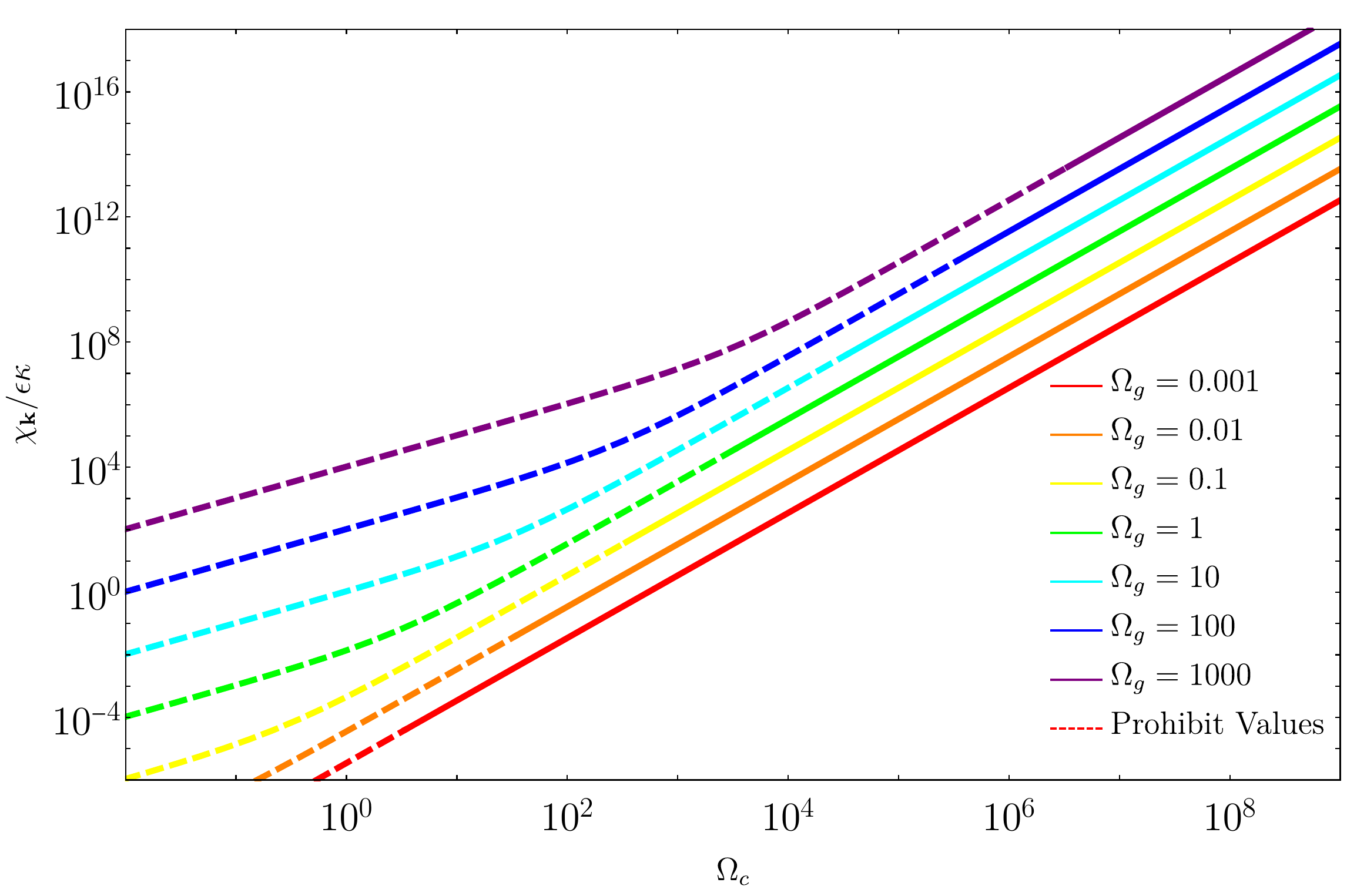}
    } \hfill
    \subfigure[$\omega_{\mathbf{k}}+\omega_{\mathbf{j}}=\Omega_c-\Omega_g$]{
        \includegraphics[width=0.475\linewidth]{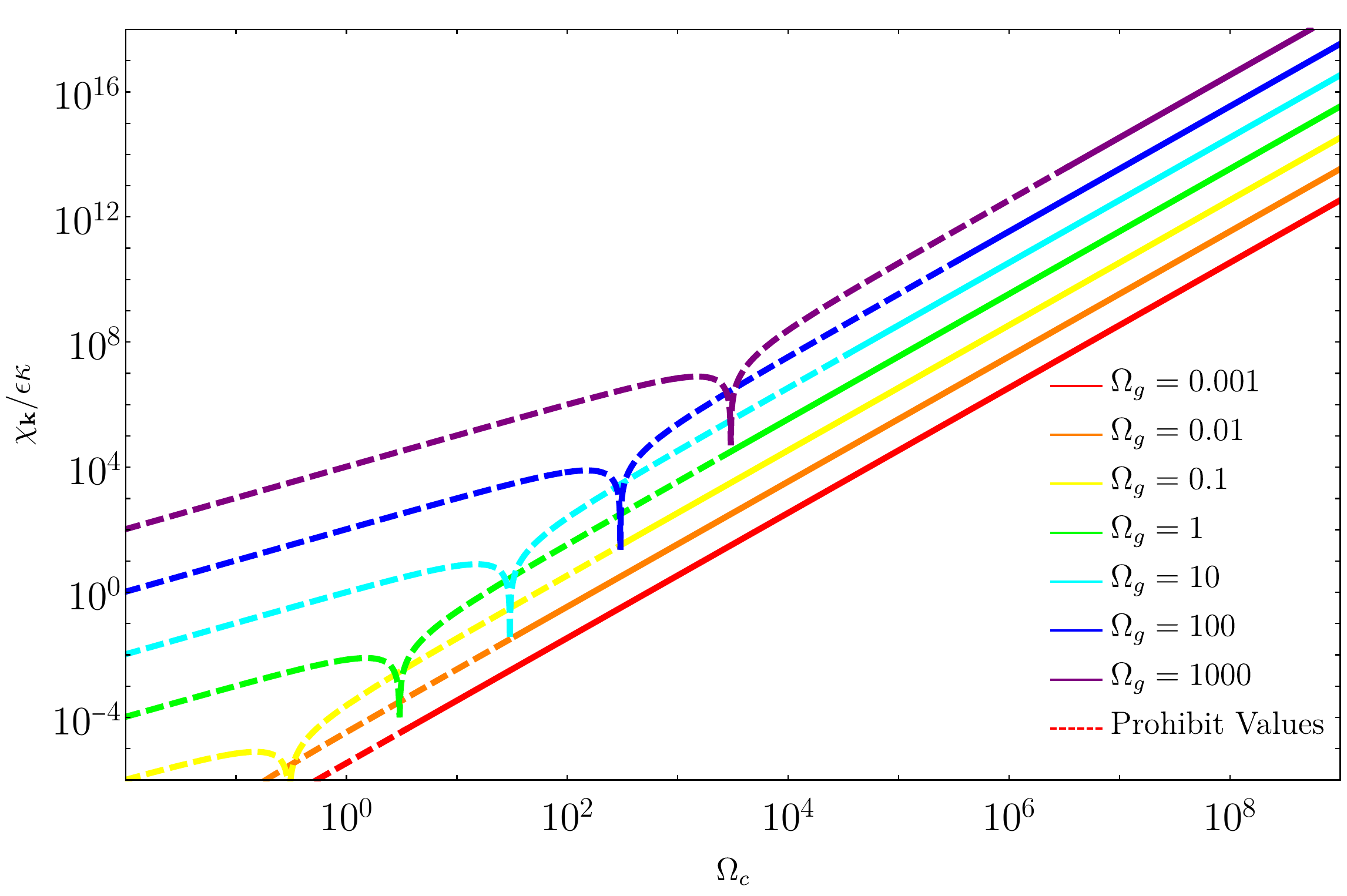}
    }
    
    \caption{\textbf{Effective amplification rate}. The figure shows coefficient $\chi_{\mathbf{k}}/\epsilon \kappa$ as a function of the cavity frequency $\Omega_c$ for different values of $\Omega_g$ (assuming $h=\kappa \Omega_g^2$, as typically expected for an oscillating binary system). We considered a cubic cavity of length $L$ and a fixed pair of modes $\mathbf{k}=(2,1,2)$ and $\mathbf{j}=(2,1,1)$, computing $\chi_{\mathbf{k}}/\epsilon \kappa$ solely as a function of $\Omega_c$. This was done by expressing the unperturbed field frequencies in terms of the cavity length $L(\Omega_c)$ required to satisfy the resonance conditions. The dashed lines indicate the values of $\Omega_c$ for which the corresponding cavity length violates the approximation $L_z \Omega_g \ll 1$ (here we imposed $L_z \Omega_g < 10^{-3}$ for graphical purposes), meaning that our analytical formulas cease to reliably describe the physics of the system in that regime. We emphasize that these results isolate the purely gravitational contribution. In a realistic experimental setup, the standard dynamical Casimir effect (driven by mechanical mirror motion) would dominate the total particle production rate. However, as discussed in the main text, the gravitational contribution possesses a distinct spectral signature which allows, at least in principle, to be distinguished from the mechanical input. Finally, this figure is intended to illustrate the qualitative behaviour, parametric dependencies, and resonant structure of the gravitational interaction in an idealized regime.}
    \label{fig:AmpVsFreq}
\end{figure*}

To illustrate the specific dependence of expressions~\eqref{eq:NkT1},~\eqref{eq:NkT2}, and~$\eqref{eq:NkT3}$ on the frequencies $\Omega_c$, $\Omega_g$, and $\omega_{\mathbf{k}}$, Figure~\ref{fig:AmpVsFreq} shows the coefficient $\chi_{\mathbf{k}}/\epsilon\kappa$, which corresponds to the effective amplification rate, as a function of the cavity frequency $\Omega_c$ for each resonance condition in Eqs.~\eqref{eq:NkT}. Due to the subtleties involved in the nontrivial relationship between the quantities $\Omega_c$ and $\omega_{\mathbf{k}}$ and the length of the cavity, it is necessary to be careful when interpreting the information presented by these graphs. This occurs because an ideal cavity can only support resonant field modes whose frequencies are inversely proportional to their size. Consequently, for a fixed gravitational wave frequency, the cavity size must be appropriately tuned to satisfy resonance conditions.

In our analysis, we started from the assumption that the gravitational wave oscillation frequency $ \Omega_g $ is externally given and, thus, not subject to experimental control. On the other hand, we considered that both the oscillation frequency of the cavity $ \Omega_c $ and the length of the unperturbed cavity, $L$, can be conveniently varied (we choose a cubic cavity $L_z=L_x=L_y=L$). 

In the graph, we choose to fix the first two subsequent modes supported by the cavity, $\mathbf{k} = (2,1,2) $ and $ \mathbf{j} = (2,1,1) $. As a consequence of fixing these modes, for each value of $\Omega_c$, the resonance condition imposes a specific value for the cavity size $L_z$ since both $\omega_{\mathbf{k},0}$ and $\omega_{\mathbf{j},0}$ depend on $L_z$. In other words, in order to maintain the resonance condition, it is necessary to change the cavity length $L_z$ as the oscillation frequency $\Omega_c$ is varied. 

However, it is important to note that our theoretical framework is based on the approximation $L_z \, \Omega_g \ll 1$, which ensures that the gravitational wave does not vary significantly across the cavity. Therefore, for values of $\Omega_c$ that lead to cavity sizes that violate this condition, the approximations underlying our model break down. In the graph, the region where this occurs is indicated by a dashed line, signalling that the corresponding values of $\Omega_c$ lie outside the validity regime of our results. Thus, while the full curve is shown for completeness, only the solid portion should be interpreted as physically reliable within the assumptions of our model.

Another important consideration in the above graphical representation is the assumption that the gravitational wave amplitude is proportional to the square of the characteristic angular frequency of the source, i.e. $h=\kappa \Omega_g^2$, where $\kappa$ is a proportionality constant that depends on the specific parameters of the source and its distance from the observer. This scaling behaviour is motivated by the quadrupole formula for gravitational radiation, which predicts that the amplitude of gravitational waves emitted by a time-varying mass quadrupole is proportional to the second time derivative of the quadrupole moment.  

Moreover, all the graphs in Fig.~\eqref{fig:AmpVsFreq} share a universal tendency where the curves corresponding to the higher oscillating frequencies of the gravitational wave $\Omega_g$ attain larger values for $\chi_{\mathbf{k}}/\epsilon\kappa$ compared to those of lower frequencies. This behaviour is physically justified by the fact that gravitational waves with higher frequencies are generally associated with higher temporal variations in the spacetime metric. Consequently, as energy can be transferred more efficiently from the gravitational wave to the quantum field, one expects an enhancement in the amplification of vacuum fluctuations into particle pairs. 

\section{Conclusions}
\label{sec:conclusions}

In this work, we investigate the dynamical Casimir effect for a scalar quantum field trapped in a three-dimensional cavity with one of its mirrors allowed to move. The entire system is subjected to a plane gravitational wave. Our main findings and conclusions are summarised as follows:
\begin{itemize}
    \item  We demonstrated that the gravitational wave modifies the proper length of the cavity, leading to a modulation of the field's natural frequencies. This interaction introduces new resonance conditions for particle creation that are distinct from those of the standard (purely mechanical) dynamical Casimir effect.
    \item  We derived analytical expressions for the particle production rates under these new resonance conditions (Eqs.~\eqref{eq:NkT1}--\eqref{eq:NkT4}). Specifically, we identified resonances at sideband frequencies $\omega_{\mathbf{k}} = |\Omega_c \pm \Omega_g|/2$, where the effective coupling strength scales with the gravitational strain amplitude $h_+$.
\end{itemize}

From an experimental standpoint, the detection of particle creation as predicted by the DCE constitutes an extremely challenging task, even without considering gravitational wave influences. This is illustrated by the fact that in more than fifty years since the seminal work of G. Moore, a concrete experimental realisation of the phenomenon has been reported only through the work of Wilson et al. in 2011~\cite{Wilson2011} and that of Lähteenmäki et al. in 2013~\cite{Lahteenmaki2013}, where effective moving boundaries were simulated by superconducting circuits in the microwave regime.  

In this respect, the prospect of detecting contributions to particle creation with an exclusive classical gravitational wave origin represents an even greater challenge. The main obstacle for detection lies in the extremely small magnitude of the gravitational wave amplitudes that reach terrestrial or near-Earth laboratories. 

Another obstacle in this task is related to the requirement for extremely high levels of precision in maintaining the resonance conditions between the unperturbed field frequencies $\omega_{\mathbf{k},0}$ and $\omega_{\mathbf{j},0}$, and both the cavity's mechanical and gravitational oscillation frequencies $\Omega_c$ and $\Omega_g$. Since the experimentalist cannot control $\Omega_g$, which solely depends on the astrophysical source, gravitationally-induced DCE must then rely on fine-tuning the cavity's parameters, such as $L_z$ and $\Omega_c$, in order to satisfy such requirements.

An additional source of complexity arises from the fact that any attempt to measure the created particles must distinguish their signatures from competing noise sources and from particles generated by purely mechanical oscillations of the cavity (in the absence of gravitational waves). Assuming that environmental noise induces a frequency shift $\Omega_c \to \Omega_c + \delta$ in the standard DCE contribution (Eq.~\eqref{eq:NkT}), the particle growth rate should maintain the amplitude scaling $\chi_{\mathbf{k}} \propto \epsilon \Omega_c$, resulting only in a shift of the spectral peak's location. In contrast, the gravitational interaction (Eq.~\eqref{eq:NkT2}) under the condition $\omega_{\mathbf{k}} = |\Omega_c \pm \Omega_g|/2$ scales as $\chi_{\mathbf{k}} \propto \epsilon h_+ (\Omega_c/\Omega_g)$. If we assume that the gravitational frequency behaves like a mechanical drift ($\Omega_g \to \delta$), the gravitational particle production shows an enhancement proportional to $\Omega_c/\delta$. Thus, even if a random shift $\delta$ in the cavity frequency mimics the frequency scale of a gravitational wave ($\delta \sim \Omega_g$), the gravitational signal should, at least in principle, be distinguishable from environmental noise due to its unique scaling signature.

From a fundamental point of view, it would be interesting to investigate the action of the quantum gravitational field on the dynamical Casimir effect. At least in the weak field limit considered here, this should be possible. This would certainly introduce decoherence due to the quantum fluctuations of space-time~\cite{Moreira2024,Moreira2025}, but it could lead to interesting new effects.

\begin{acknowledgments}
The authors acknowledge CNPq through grant 308065/2022-0, the financial support of the National Institute of Science and Technology for Applied Quantum Computing through CNPq grant 408884/2024-0, the Coordination of Superior Level Staff Improvement (CAPES), and FAPEG through grant 202510267001843.
\end{acknowledgments}

\appendix

\section{Instantaneous mode function}
\label{AppendixA}

By considering high frequency waves such that $\Bar{h}_{ij}(\mathbf{x},t)$ does not vary appreciably with position inside the cuboid, we write
\begin{equation}
    h_s\qty[\omega_{\mathbf{k}}^{(1)}(t)]^2=-\Bar{h}_{ij}(t)\int_{\Sigma(t)}\dd^3x\,\varphi_{\mathbf{k}}^{(0)}(\mathbf{x},t)\partial^i\partial^j\varphi_{\mathbf{k}}^{(0)}(\mathbf{x},t). 
\end{equation}
Let us now note that terms with $i\neq j$ will lead to integrals of the form
\begin{equation}
    \int_0^{L}\dd\xi\,\sin(k_\xi\xi)\cos(k_\xi\xi)=0.
\end{equation}
Therefore, only diagonal terms in the metric perturbation field will contribute. In terms of the gravitational wave travelling in the $z-$direction, Eq.~\eqref{Gravitational-wave-z-direction}, we see that only the plus polarisation leads to a non-zero contribution, $h_s=h_+$. We may then write
\begin{eqnarray}
\label{First-order-correction-frequency}
    \qty[\omega_{\mathbf{k}}^{(1)}(t)]^2&=&-\cos(\Omega_gt+\delta_g)\int_{\Sigma(t)}\dd^3x\,\varphi_{\mathbf{k}}^{(0)}(\mathbf{x},t)\qty(\partial_x^2-\partial_y^2)\varphi_{\mathbf{k}}^{(0)}(\mathbf{x},t) \nonumber \\
    &=&\qty(k_x^2-k_y^2)\cos(\Omega_gt+\delta_g)
\end{eqnarray}

Lastly, putting everything together, we obtain the following
\begin{equation}
    \nabla^2\varphi_{\mathbf{k}}^{(1)}(\mathbf{x},t)=-\qty[\omega_{\mathbf{k}}^{(0)}(t)]^2\varphi_{\mathbf{k}}^{(1)}(\mathbf{x},t)
\end{equation}
for the first order correction to the mode functions. This implies either $\varphi_{\mathbf{k}}^{(1)}(\mathbf{x},t)=0$ or $\varphi_{\mathbf{k}}^{(1)}(\mathbf{x},t)=\varphi_{\mathbf{k}}^{(0)}(\mathbf{x},t)$. However, the latter cannot hold due to the orthogonality condition of the mode functions, so we must have the following
\begin{equation}
    \varphi_{\mathbf{k}}^{(1)}(\mathbf{x},t)=0.
\end{equation}

In summary, one has
\begin{subequations}
\begin{equation}
    \varphi_{\mathbf{k}}(\mathbf{x},t)=\varphi_{\mathbf{k}}^{(0)}(\mathbf{x},t)+O(h^2),
\end{equation}
\begin{equation} \label{eq:timefreqApp}
    \omega_{\mathbf{k}}^2(t)=[\omega_{\mathbf{k}}^{(0)}(t)]^2+h_+(k_x^2-k_y^2)\cos(\Omega_g t)+O(h^2),
\end{equation}
\end{subequations}
where we dropped the phase factor $\delta_g$ for convenience.

\section{Hamiltonian derivations}
\label{AppendixB}

We start by deriving the expressions for the ODEs~\eqref{eq:dqdp}, which describe the time derivatives for the position and momentum quadrature operators $\hat{q}_{\mathbf{k}}(t)$ and $\hat{p}_{\mathbf{k}}(t)$. 

\subsection{Time derivatives of the quadrature operators}
From the $\hat{q}_{\mathbf{k}}$ and $\hat{p}_{\mathbf{k}}$ definitions~\eqref{eq:qp} together with the expressions~\eqref{eq:dyneqn} and~\eqref{fieldsexp}, one finds
\begin{eqnarray}
    \nonumber\frac{\dd \hat{q}_{\mathbf{k}}}{\dd t}&=&\int_{\Sigma(t)}\dd^3 x~\left[\partial_t\hat{\Phi}\varphi_{\mathbf{k}}+\hat{\Phi}\dot{\varphi}_{\mathbf{k}}\right] =\int_{\Sigma(t)}\dd^3 x~\left[\hat{\Pi}\varphi_{\mathbf{k}}+\hat{\Phi}\dot{\varphi}_{\mathbf{k}}\right]\nonumber\\
    &=&\sum_{\mathbf{j}}\hat{p}_{\mathbf{j}}\int_{\Sigma(t)}\dd^3 x~\varphi_{\mathbf{j}}\varphi_{\mathbf{k}}+\sum_{\mathbf{j}}\hat{q}_{\mathbf{j}}\int_{\Sigma(t)}\dd^3 x~\varphi_{\mathbf{j}}\dot{\varphi}_{\mathbf{k}}\nonumber\\
    &=&\hat{p}_{\mathbf{k}}+\sum_{\mathbf{j}}G_{\mathbf{k},\mathbf{j}}(t)\hat{q}_{\mathbf{j}},
\end{eqnarray}
and
\begin{eqnarray}
    \nonumber\frac{\dd \hat{p}_{\mathbf{k}}}{\dd t}&=&\int_{\Sigma(t)}\dd^3 x~\left[\partial_t\hat{\Pi}\varphi_{\mathbf{k}}+\hat{\Pi}\dot{\varphi}_{\mathbf{k}}\right]
    =\int_{\Sigma}\dd^3 x~\left[\nabla^2 \hat{\Phi}\varphi_{\mathbf{k}}+\Bar{h}_{ij}\partial^i\partial^j\hat{\Phi}\varphi_{\mathbf{k}}+\hat{\Pi}\dot{\varphi}_{\mathbf{k}}\right]\nonumber\\
    &=&\sum_{\mathbf{j}}\int_{\Sigma}\dd^3 x~\left[ \hat{q}_{\mathbf{j}}(\nabla^2+\Bar{h}_{ij}\partial^i\partial^j)\varphi_{\mathbf{j}}\varphi_{\mathbf{k}}+\hat{p}_{\mathbf{j}}\varphi_{\mathbf{j}}\dot{\varphi}_{\mathbf{k}}\right] \nonumber\\
    &=&-\omega_{\mathbf{k}}^2(t)\hat{q}_{\mathbf{k}}-\sum_{\mathbf{j}}G_{\mathbf{j},\mathbf{k}}(t)\hat{p}_{\mathbf{j}}.
\end{eqnarray}

\subsection{Time derivatives of the annihilation and creation operators}

Next, we find the ODEs~\eqref{eq:odeaad} for the instantaneous creation and annihilation operators $\hat{a}_{\mathbf{k}}(t)$ and $\hat{a}_{\mathbf{k}}^{\dagger}(t)$. Taking into account the inverses of Eqs.~\eqref{eq:aad}, we obtain
\begin{equation}
        \hat{p}_{\mathbf{k}}=i\sqrt{\frac{\omega_{\mathbf{k}}}{2}}\left(a_{\mathbf{k}}^{\dagger}e^{i\Theta_{\mathbf{k}}}-a_{\mathbf{k}}e^{-i\Theta_{\mathbf{k}}}\right),\qquad
        \hat{q}_{\mathbf{k}}=\frac{1}{\sqrt{2\omega_{\mathbf{k}}}}\left(a_{\mathbf{k}}^{\dagger}e^{i\Theta_{\mathbf{k}}}+a_{\mathbf{k}}e^{-i\Theta_{\mathbf{k}}}\right).
\end{equation}
Thus
\begin{eqnarray}
    \nonumber\frac{d\hat{p}_{\mathbf{k}}}{dt}&=&-\omega_{\mathbf{k}}^2\hat{q}_{\mathbf{k}}-\sum_{\mathbf{j}} G_{\mathbf{j},\mathbf{k}}\hat{p}_\mathbf{j}\nonumber\\
    &=&-\omega_{\mathbf{k}}\sqrt{\frac{\omega_{\mathbf{k}}}{2}}\left(a_{\mathbf{k}}^{\dagger}e^{i\Theta_{\mathbf{k}}}+a_{\mathbf{k}}e^{-i\Theta_{\mathbf{k}}}\right)
    -i\sum_{\mathbf{j}} G_{\mathbf{j},\mathbf{k}}\sqrt{\frac{\omega_{\mathbf{j}}}{2}}\left(a_{\mathbf{j}}^{\dagger}e^{i\Theta_{\mathbf{j}}}-a_{\mathbf{j}}e^{-i\Theta_{\mathbf{j}}}\right)
\end{eqnarray}
and
\begin{eqnarray}
    \frac{\dd\hat{q}_{\mathbf{k}}}{\dd t}\nonumber&=&\hat{p}_{\mathbf{k}}+\sum_{\mathbf{j}} G_{\mathbf{k},\mathbf{j}}\hat{q}_{\mathbf{j}}\nonumber\\
    &=&i\sqrt{\frac{\omega_{\mathbf{k}}}{2}}\left(a_{\mathbf{k}}^{\dagger}e^{i\Theta_{\mathbf{k}}}-a_{\mathbf{k}}e^{-i\Theta_{\mathbf{k}}}\right)
    +\sum_{\mathbf{j}} G_{\mathbf{k},\mathbf{j}}\frac{1}{\sqrt{2\omega_{\mathbf{j}}}}\left(a_{\mathbf{j}}^{\dagger}e^{i\Theta_{\mathbf{j}}}+a_{\mathbf{j}}e^{-i\Theta_{\mathbf{j}}}\right).
\end{eqnarray}

Taking the time derivatives from definitions~\eqref{eq:aad}:
\begin{eqnarray}
    \dot{\hat{a}}_{\mathbf{k}}e^{-i\Theta_{\mathbf{k}}}&=&-\frac{1}{2}\frac{\dot{\omega}_{\mathbf{k}}}{\omega_{\mathbf{k}}}\frac{1}{\sqrt{2\omega_{\mathbf{k}}}}\left[\omega_{\mathbf{k}}\hat{q}_{\mathbf{k}}+i\hat{p}_{\mathbf{k}}\right]
   +\frac{1}{\sqrt{2\omega_{\mathbf{k}}}}\left[\dot{\omega}_{\mathbf{k}}\hat{q}_ {\mathbf{k}}+\omega_{\mathbf{k}}\dot{\hat{q}}_{\mathbf{k}}+i\dot{\hat{p}}_{\mathbf{k}}\right]+i\omega_{\mathbf{k}}\hat{a}_{\mathbf{k}}e^{-i\Theta_{\mathbf{k}}}\nonumber\\
    &=&\mu_{\mathbf{k},\mathbf{k}}\hat{a}_{\mathbf{k}}^{\dagger}e^{i\Theta_{\mathbf{k}}}+\frac{1}{2}\sum_{\mathbf{j}}[\mu_{\mathbf{k},\mathbf{j}}+\mu_{\mathbf{j},\mathbf{k}}]\hat{a}_{\mathbf{j}}^{\dagger}e^{i\Theta_{\mathbf{j}}}
    +\frac{1}{2}\sum_{\mathbf{j}}[\mu_{\mathbf{k},\mathbf{j}}-\mu_{\mathbf{j},\mathbf{k}}]\hat{a}_{\mathbf{j}}e^{-i\Theta_{\mathbf{j}}},
\end{eqnarray}
allowing thus to write
\begin{equation}
    \frac{\dd \hat{a}_{\mathbf{k}}}{\dd t}=\sum_{\mathbf{k}}\left[\mu_{(\mathbf{k},\mathbf{j})}\hat{a}_{\mathbf{j}}^{\dagger}e^{i[\Theta_{\mathbf{k}}+\Theta_{\mathbf{j}}]}+\mu_{[\mathbf{k},\mathbf{j}]}\hat{a}_{\mathbf{j}}e^{-i[\Theta_{\mathbf{k}}-\Theta_{\mathbf{j}}]}\right].
\end{equation}

\subsection{Effective Hamiltonian}

With the ingredients derived in the last section we can construct the effective Hamiltonian that generates the dynamical equations~\eqref{eq:odeaad} by considering the most general quadratic operator
\begin{equation}
    \label{ht}\hat{H}(t)=\sum_{kl}\left[\mathcal{A}_{kl}(t)\hat{a}_k^{\dagger}(t)\hat{a}_l^{\dagger}(t)+\mathcal{B}_{kl}(t)\hat{a}_k^{\dagger}(t)\hat{a}_l(t)+\mathcal{C}_{kl}(t)\hat{a}_l^{\dagger}(t)\hat{a}_k(t)+\mathcal{D}_{kl}(t)\hat{a}_k(t)\hat{a}_l(t)\right],
\end{equation}
which is: (i) hermitian, by satisfying the conditions $\mathcal{A}_{kl}(t)=\mathcal{D}_{kl}^*(t)$, $\mathcal{B}_{kl}(t)=\mathcal{C}_{kl}^*(t)$ and (ii) invariant over an index change, with conditions $\mathcal{A}_{kl}(t)=\mathcal{A}_{lk}(t)$, $\mathcal{D}_{kl}(t)=\mathcal{D}_{lk}(t)$, $\mathcal{B}_{kl}(t)=\mathcal{C}_{lk}(t)$ and $\mathcal{B}_{lk}(t)=\mathcal{C}_{kl}(t)$.

Suppressing the notation for time dependence, the correspondent Heisenberg equation of motion for the annihilation and creation operators is therefore
\begin{equation}
    \dot{\hat{a}}_j =i\left[\hat{H},\hat{a}_j\right]
    =-i\sum_k\bigg[\left(\mathcal{A}_{kj}+\mathcal{A}_{jk}\right)\hat{a}_k^{\dagger}+\left(\mathcal{B}_{jk}+\mathcal{C}_{kj}\right)\hat{a}_k\bigg]
    \label{daH}
\end{equation}
and
\begin{equation}
    \dot{\hat{a}}_j^{\dagger} =i\left[\hat{H},\hat{a}_j^{\dagger}\right]
    =i\sum_k\bigg[\left(\mathcal{D}_{kj}+\mathcal{D}_{jk}\right)\hat{a}_k+\left(\mathcal{B}_{kj}+\mathcal{C}_{jk}\right)\hat{a}_k^{\dagger}\bigg].
    \label{dadH}
\end{equation}

Comparing~\eqref{da} with~\eqref{daH} and~\eqref{dad} with~\eqref{dadH}, we obtain the following system
\begin{subequations}
\label{coef}
    \begin{align*}
    -i\left[\mathcal{A}_{kj}(t)+\mathcal{A}_{jk}(t)\right]&=-2i\mathcal{A}_{kj}(t)=B_{kj}^*(t),\\
    -i\left[\mathcal{C}_{kj}(t)+\mathcal{B}_{jk}(t)\right]&=-2i\mathcal{C}_{kj}(t)=A_{kj}(t),\\
    i\left[\mathcal{D}_{kj}(t)+\mathcal{D}_{jk}(t)\right]&=2i\mathcal{D}_{kj}(t)=B_{kj}(t),\\
    i\left[\mathcal{B}_{kj}(t)+\mathcal{C}_{jk}(t)\right]&=2i\mathcal{B}_{kj}(t)=A_{kj}^*(t).
\end{align*}
\end{subequations}
Inserting the last coefficients into Eq.~\eqref{ht}, one obtains the following expression for the effective Hamiltonian in the Schrodinger picture
\begin{align}
\label{HS}
\hat{H}_S(t) &=\frac{i}{2}\sum_{{j} {k}} \Bigg[A_{kj}(t)\hat{a}_{j}^{\dagger}\hat{a}_{k}+B_{kj}^*(t)\hat{a}_{j}^{\dagger}\hat{a}_{k}^{\dagger}-\text{h.c.}
\Bigg].
\end{align}

\section{Obtaining the Bogoliubov coefficients}
\label{AppendixBogoliubov}
From the differential equation for $\hat{a}_{\mathbf{j}}$ and $ \hat{a}_{\mathbf{j}}^{\dagger}$ as
\begin{subequations}
    \label{aaddynmeqn}
    \begin{align}
   \label{da} \dot{\hat{a}}_{\mathbf{j}}&=\sum_{\mathbf{k}}\left(A_{\mathbf{k}\mathbf{j}}\hat{a}_{\mathbf{k}}+B_{\mathbf{k}\mathbf{j}}^*\hat{a}_{\mathbf{k}}^{\dagger}\right),\\
   \label{dad} \dot{\hat{a}}_{\mathbf{j}}^{\dagger}&=\sum_{\mathbf{k}}\left(B_{\mathbf{k}\mathbf{j}}\hat{a}_{\mathbf{k}}+A_{\mathbf{k}\mathbf{j}}^*\hat{a}_{\mathbf{k}}^{\dagger}\right),
\end{align}
\end{subequations}
where
\begin{equation}
    A_{\mathbf{k}\mathbf{j}}=\frac{1}{2}\left[\mu_{\mathbf{k}\mathbf{j}}-\mu_{\mathbf{j}\mathbf{k}}\right]e^{-i\left[\Theta_{\mathbf{k}}-\Theta_{\mathbf{j}}\right]}\qquad
    B_{\mathbf{k}\mathbf{j}}=\frac{1}{2}\left[\mu_{\mathbf{k}\mathbf{j}}+\mu_{\mathbf{j}\mathbf{k}}\right]e^{-i\left[\Theta_{\mathbf{k}}+\Theta_{\mathbf{j}}\right]}.
\end{equation}

Solutions for the set of differential equations~\eqref{aaddynmeqn} can be obtained through considerations of the Bogoliubov transformations 
\begin{align}
\label{akab}
    \hat{a}_{\mathbf{k}}(t)&=\sum_{\mathbf{j}}\left[\alpha_{\mathbf{j}\mathbf{k}}(t)\hat{a}_{\mathbf{j}}(0)+\beta_{\mathbf{j}\mathbf{k}}^*(t)\hat{a}_{\mathbf{j}}^{\dagger}(0)\right]
\end{align}
where $\alpha_{\mathbf{j}\mathbf{k}}(t)$ and $\beta_{\mathbf{j}\mathbf{k}}(t)$ are the "instantaneous" Bogoliubov coefficients with the initial conditions $\alpha_{\mathbf{j}\mathbf{k}}(0)=\delta_{\mathbf{j}\mathbf{k}}$ and $\beta_{\mathbf{j}\mathbf{k}}(0)=0$. Inserting the expression~\eqref{akab} and its hermitian conjugated into Eq.~\eqref{da} one obtains
\begin{equation}
\label{aaddynmeqn}
    \dot{\hat{a}}_{\mathbf{k}}(t)=\sum_{\mathbf{j}\mathbf{j^{\prime}}}\bigg\{\left[A_{\mathbf{k}\mathbf{j}}(t)\alpha_{\mathbf{j'}\mathbf{j}}(t)+B_{\mathbf{k}\mathbf{j}}^*(t)\beta_{\mathbf{j'}\mathbf{j}}(t)\right]\hat{a}_{\mathbf{j}^{\prime}}(0)
    +\left[B_{\mathbf{k}\mathbf{j}}^*(t)\alpha_{\mathbf{j'}\mathbf{j}}^*(t)+A_{\mathbf{k}\mathbf{j}}(t)\beta_{\mathbf{j'}\mathbf{j}}^*(t)\right]\hat{a}_{\mathbf{j}^{\prime}}^{\dagger}(0)\bigg\}.
\end{equation}

Equating the last expression with the time derivative of eq.~\eqref{akab} given by $\dot{\hat{a}}_{\mathbf{k}}(t)=\sum_{\mathbf{j}}\left[\dot{\alpha}_{\mathbf{j}\mathbf{k}}(t)\hat{a}_{\mathbf{j}}(0)+\dot{\beta}_{\mathbf{j}\mathbf{k}}^*(t)\hat{a}_{\mathbf{j}}^{\dagger}(0)\right]$, we find the following differential equations for the Bogoliubov coefficients
\begin{subequations}
    \label{debog}
    \begin{align}
        \dot{\alpha}_{\mathbf{j}\mathbf{k}}(t)&=\sum_{\mathbf{j}^{\prime}}\left[A_{\mathbf{k}\mathbf{j}}(t)\alpha_{\mathbf{j'}\mathbf{j}}(t)+B_{\mathbf{k}\mathbf{j}}^*(t)\beta_{\mathbf{j'}\mathbf{j}}(t)\right],\\
        \dot{\beta}_{\mathbf{j}\mathbf{k}}(t)&=\sum_{\mathbf{j}^{\prime}}\left[B_{\mathbf{k}\mathbf{j}}(t)\alpha_{\mathbf{j'}\mathbf{j}}(t)+A_{\mathbf{k}\mathbf{j}}^*(t)\beta_{\mathbf{j'}\mathbf{j}}(t)\right].
    \end{align}
\end{subequations}
Since the coefficients $A_{\mathbf{k}\mathbf{j}}(t)$ and $B_{\mathbf{k}\mathbf{j}}(t)$ have a dependence in first order in $\epsilon$ and $ h $, if we suppose the cavity to move much slower than the speed of light ($\dot{l}(t)\ll 1$), one can expand the Bogoliubov coefficients in terms of $\epsilon$ and $ h $. The zero-th order solution is
\begin{equation}
\label{zerobogo}
    \alpha^{(0)}_{\mathbf{j}\mathbf{k}}(t)=\alpha_{\mathbf{j}\mathbf{k}}(0)= \delta_{\mathbf{j}\mathbf{k}}, \qquad
     \beta^{(0)}_{\mathbf{j}\mathbf{k}}(t)=\beta_{\mathbf{j}\mathbf{k}}(0)=0.
\end{equation}
Substituting Eq.~\eqref{zerobogo} into the right side of Eq.~\eqref{zerobogo} we obtain the following
\begin{subequations}
    \label{debog2}
    \begin{align}
        \dot{\alpha}_{\mathbf{j}\mathbf{k}}^{(1)}(t)&=\sum_{\mathbf{j}^{\prime}}\left[A_{\mathbf{k}\mathbf{j}}(t)\alpha_{\mathbf{j}\mathbf{j^{\prime}}}^{(0)}(t)+B_{\mathbf{k}\mathbf{j}}^*(t)\beta_{\mathbf{j}\mathbf{j^{\prime}}}^{(0)}(t)\right],\\
        \dot{\beta}_{\mathbf{j}\mathbf{k}}^{(1)}(t)&=\sum_{\mathbf{j}^{\prime}}\left[B_{\mathbf{k}\mathbf{j}}(t)\alpha_{\mathbf{j}\mathbf{j^{\prime}}}^{(0)}(t)+A_{\mathbf{k}\mathbf{j}}^*(t)\beta_{\mathbf{j}\mathbf{j^{\prime}}}^{(0)}(t)\right],
    \end{align}
\end{subequations}
whose solutions are given by
\begin{equation}
\label{alphabetafirstorder}
    \alpha_{\mathbf{j}\mathbf{k}}^{(1)}(t)=\int_0^t dt^{\prime}~A_{\mathbf{k}\mathbf{j}}(t^{\prime}),\qquad
    \beta_{\mathbf{j}\mathbf{k}}^{(1)}(t)=\int_0^t dt^{\prime}~B_{\mathbf{k}\mathbf{j}}(t^{\prime}).
\end{equation}
Continuing with the same iteration scheme, solutions can be obtained at high orders of magnitude of the perturbation parameter
\begin{subequations}
    \label{debog2}
    \begin{align}
        \dot{\alpha}_{\mathbf{j}\mathbf{k}}^{(2)}(t)&=\sum_{\mathbf{j}^{\prime}}\left[A_{\mathbf{k}\mathbf{j}}(t)\alpha_{\mathbf{j}\mathbf{j^{\prime}}}^{(1)}(t)+B_{\mathbf{k}\mathbf{j}}^*(t)\beta_{\mathbf{j}\mathbf{j^{\prime}}}^{(1)}(t)\right],\\
        \dot{\beta}_{\mathbf{j}\mathbf{k}}^{(2)}(t)&=\sum_{\mathbf{j}^{\prime}}\left[B_{\mathbf{k}\mathbf{j}}(t)\alpha_{\mathbf{j}\mathbf{j^{\prime}}}^{(1)}(t)+A_{\mathbf{k}\mathbf{j}}^*(t)\beta_{\mathbf{j}\mathbf{j^{\prime}}}^{(1)}(t)\right],
    \end{align}
\end{subequations}
and consequently
\begin{align}
       \alpha_{\mathbf{j}\mathbf{k}}^{(2)}(t)&=\frac{1}{2}\sum_{\mathbf{j}^{\prime}}\left[\alpha^{(1)}_{\mathbf{k}\mathbf{j}}(t)\alpha_{\mathbf{j}\mathbf{j^{\prime}}}^{(1)}(t)+\beta^{(1)*}_{\mathbf{k}\mathbf{j}}(t)\beta_{\mathbf{j}\mathbf{j^{\prime}}}^{(1)}(t)\right],\\
        \beta_{\mathbf{j}\mathbf{k}}^{(2)}(t)&=\frac{1}{2}\sum_{\mathbf{j}^{\prime}}\left[\beta^{(1)}_{\mathbf{k}\mathbf{j}}(t)\alpha_{\mathbf{j}\mathbf{j^{\prime}}}^{(1)}(t)+\alpha^{(1)*}_{\mathbf{k}\mathbf{j}}(t)\beta^{(1)}_{\mathbf{j}\mathbf{j^{\prime}}}(t)\right],    
\end{align}
As we will only be interested in the first order solution for the Bogoliubov coefficients, we resume the Bogoliubov expressions to be 
\begin{equation}
    \alpha_{\mathbf{j}\mathbf{k}}(t)=\delta_{\mathbf{j}\mathbf{k}}+\int_0^t \dd t^{\prime}~A_{\mathbf{k}\mathbf{j}}(t^{\prime}),\qquad
    \beta_{\mathbf{j}\mathbf{k}}(t)=\int_0^t \dd t^{\prime}~B_{\mathbf{k}\mathbf{j}}(t^{\prime}).
\end{equation}

\section{Calculations}
\label{AppendixCalc}
In this appendix, we will tackle the question of how to compute the number of particles created from the vacuum using time-independent Hamiltonians in the resonance regime. 

\subsection{Calculating the number of particles}
\label{AppendixC}

Under the assumption that the cavity setup is able to couple at most two field modes with each other, together with the rotating-wave approximation, one expects to simplify the effective Hamiltonian~\eqref{eq:effham} to the form 
\begin{equation}
    \hat{H}=\frac{i}{2}\left[g\hat{a}_{\mathbf{k}}^{\dagger}\hat{a}_{\mathbf{p}}^{\dagger}-g^*\hat{a}_{\mathbf{k}}\hat{a}_{\mathbf{p}}\right],
\end{equation}
where $g=|g|e^{i\delta}$ is a complex coefficient. In this special case, the time evolution for the annihilation operator takes the closed form for $\mathbf{k}=\mathbf{p}$, 
\begin{align}
\nonumber\hat{a}_{\mathbf{k}}(t)&=e^{i\hat{H}t}\hat{a}_{\mathbf{k}}e^{-i\hat{H}t}=\left(1+\frac{(gt)^2}{2!}+\frac{(gt)^4}{4!}+\dots\right)\hat{a}_{\mathbf{k}}
+\left(gt+\frac{(gt)^3}{3!}+\frac{(gt)^5}{5!}\dots\right)\hat{a}_{\mathbf{k}}^{\dagger} \nonumber\\
&=\cosh(|g|t)\hat{a}_{\mathbf{k}}+\sinh(|g|t)e^{i\delta}\hat{a}_{\mathbf{k}}^{\dagger},
\end{align}
where we have used the following identity (which can be derived from the Baker-Campbell-Hausdorff formula)
\begin{equation}
    e^{i\hat{V}\lambda}\hat{O}e^{-i\hat{V}\lambda}=\hat{O}+i\lambda\comm{\hat{V}}{\hat{O}}+\frac{(i\lambda)^2}{2!}\comm{\hat{V}}{\comm{\hat{V}}{\hat{O}}}
    +\frac{(i\lambda)^3}{3!}\comm{\hat{V}}{\comm{\hat{V}}{\comm{\hat{V}}{\hat{O}}}}+\dots.
\end{equation}
In the case of $\mathbf{k}\neq\mathbf{p}$, the instantaneous annihilation operators take the form
\begin{equation}
    \hat{a}_{\mathbf{k}}(t)=\cosh(\frac{|g|t}{2})\hat{a}_{\mathbf{k}}+\sinh(\frac{|g|t}{2})e^{i\delta}\hat{a}_{\mathbf{p}}^{\dagger}.
\end{equation}

This means that for $\hat{a}_{\mathbf{p}}\ket{0}=0$ and $\hat{a}_{\mathbf{p}}\ket{0}=\ket{1_{\mathbf{p}}}$, the average number of particles in the $\mathbf{p}$-th mode is 
\begin{equation}
\label{eq:Np}
    \langle N_{\mathbf{p}}\rangle = \bra{0}\hat{a}_{\mathbf{p}}^{\dagger}(t)\hat{a}_{\mathbf{p}}(t)\ket{0}=\sinh^2\left(\frac{|g|t}{2}\right).
\end{equation}

\subsection{Applying the rotating-wave approximation}

We search for resonance conditions that amplify the particle creation process from the vacuum from the effective Hamiltonian
\begin{equation}
    \hat{H}_{\text{eff}}(t)=\frac{i}{2}\sum_{\mathbf{k}} B_{\mathbf{k},\mathbf{k}}(t)\hat{a}_{\mathbf{k}}^{\dagger 2}
    +\frac{i}{2}\sum_{\mathbf{j}\neq\mathbf{k}} \Bigg[B_{\mathbf{k},\mathbf{j}}(t)\hat{a}_{\mathbf{j}}^{\dagger}\hat{a}_{\mathbf{k}}^{\dagger}+A_{\mathbf{k},\mathbf{j}}(t)\hat{a}_{\mathbf{j}}^{\dagger}\hat{a}_{\mathbf{k}}\Bigg]+\text{h.c.},
\end{equation}
where
\begin{align}
    B_{\mathbf{k},\mathbf{k}}(t)&=\frac{1}{2}\left(\mu_{\mathbf{k},\mathbf{j}}(t)+\mu_{\mathbf{j},\mathbf{k}}(t)\right)e^{2i\Theta_{\mathbf{k}}(t)},\\
    B_{\mathbf{k},\mathbf{j}}(t)&=\frac{1}{2}\left(\mu_{\mathbf{k},\mathbf{j}}(t)+\mu_{\mathbf{j},\mathbf{k}}(t)\right)e^{i[\Theta_{\mathbf{k}}(t)+\Theta_{\mathbf{j}}(t)]},\\
    A_{\mathbf{k},\mathbf{j}}(t)&=\frac{1}{2}\left(\mu_{\mathbf{k},\mathbf{j}}(t)+\mu_{\mathbf{j},\mathbf{k}}(t)\right)e^{-i[\Theta_{\mathbf{k}}(t)-\Theta_{\mathbf{j}}(t)]},
\end{align}
with $\mu_{\mathbf{k},\mathbf{j}}(t)$ given by Eq.~\eqref{eq:mukj}, while $\Theta(t)=\int_0^{t'}\dd t'~\omega_{\mathbf{k}}(t')$.

We assume the cavity to perform weak oscillatory motion
\begin{equation}
    L(t)=L_z[1+\epsilon \sin (\Omega_c t)],
\end{equation}
where $\epsilon \ll 1$ and $\Omega_c$ is the oscillation frequency of the cavity.

With these ingredients at hand, we begin computing a second-order expansion for the time-dependent frequency $\omega_{\mathbf{k}}$ in terms of the parameters $\epsilon$ and $h_+$, as given by 
\begin{equation}
    \label{eq:wk}
    \omega_{\mathbf{k}}(t)
    = \omega_{\mathbf{k},0}\left\{ 1-\frac{k_z^2}{\omega_{\mathbf{k},0}^2}\epsilon\sin(\Omega_c t) + \frac{h_+}{2} \frac{k_x^2-k_y^2}{\omega_{\mathbf{k},0}^2} \cos(\Omega_g t) 
    + \frac{\epsilon h_+}{2}\frac{k_z^2}{\omega_{\mathbf{k},0}^2}\frac{k_x^2-k_y^2}{\omega_{\mathbf{k},0}^2} \sin(\Omega_c t)\cos(\Omega_g t)\right\},
\end{equation}
where we have defined $\omega_{\mathbf{k},0}^2=k_x^2+k_y^2+k_z^2(0)$.

\subsubsection{For the $B_{\mathbf{k},\mathbf{k}}(t)$ terms}

Here we will analyse how to simplify the part of the Hamiltonian proportional to $B_{\mathbf{k},\mathbf{k}}(t)$ by applying the rotating-wave approximation. From Eq.~\eqref{eq:wk} one can easily obtain the second-order expansion for $\mu_{\mathbf{k},\mathbf{j}}$, 
\begin{align*}
    \mu_{\mathbf{k}\mathbf{k}}(t)=-\frac{\dot{\omega}_{\mathbf{k}}}{2\omega_{\mathbf{k}}}&=\frac{1}{2}\epsilon\Omega_c\frac{k_z^2}{\omega_{\mathbf{k},0}^2}\cos(\Omega_c t) + \frac{h_+}{4}\Omega_g \frac{k_x^2-k_y^2}{\omega_{\mathbf{k},0}^2} \sin(\Omega_g t) \nonumber\\
    &+\frac{\epsilon h_+ k_z^2}{2 }\frac{k_x^2-k_y^2}{\omega_{\mathbf{k},0}^4}\left[ \Omega_g  \sin (\Omega_c t) \sin (\Omega_g t)-\Omega_c \cos (\Omega_c t) \cos (\Omega_g t )\right],
\end{align*}
and the following expansion in order $h\epsilon$:
\begin{align}
    e^{2i\Theta_{\mathbf{k}}(t)}=\Bigg[1&-\frac{2 i  \epsilon }{\Omega_c}\frac{k_z^2}{\omega_{\mathbf{k},0} }+\frac{2i\epsilon}{\Omega_c}\frac{ k_z^2}{\omega_{\mathbf{k},0}}\cos(\Omega_c t)+\frac{ih_+}{\Omega_g}\frac{k_x^2-k_y^2}{\omega_{\mathbf{k},0}}\sin(\Omega_g t)\Bigg]e^{2i\omega_{\mathbf{k},0}t}.
\end{align}
Combining them into the expression $B_{\mathbf{k},\mathbf{k}}(t)$ one obtains at second order
\begin{eqnarray*}
 B_{\mathbf{k},\mathbf{k}}(t)&=& \Bigg[k_z^2 \epsilon \Omega_c \cos(\Omega_c t) - h_+ k_z^2 \epsilon \Omega_c\frac{k_x^2-k_y^2}{\omega_{\mathbf{k},0}^2} \cos(\Omega_c t) \cos(\Omega_g t) 
 +\frac{h_+}{2}\Omega_g(k_x^2-k_y^2) \left( 1 - \frac{2i k_z^2 \epsilon }{\omega_{\mathbf{k},0} \Omega_c} \right) \sin(\Omega_g t) \\
 &+& i\epsilon h_+k_z^2 \frac{k_x^2-k_y^2}{\omega_{\mathbf{k},0}} \left( \frac{\Omega_g}{\Omega_c} + \frac{\Omega_c}{\Omega_g} \right) \cos(\Omega_c t) \sin(\Omega_g t) 
 + \epsilon h_+ k_z^2 \Omega_g\frac{k_x^2-k_y^2}{\omega_{\mathbf{k},0}^2} \sin(\Omega_c t) \sin(\Omega_g t) \Bigg] \frac{e^{2 i \omega_{\mathbf{k},0} t}}{2 \omega_{\mathbf{k},0}^2}.
\end{eqnarray*}
By expressing all the periodic functions in terms of complex exponentials, one obtains
\begin{align}
 B_{\mathbf{k},\mathbf{k}}(t)&=\frac{1}{4}\epsilon\Omega_c\frac{k_z^2}{\omega_{\mathbf{k},0}^2}e^{i[2\omega_{\mathbf{k},0}+\Omega_c]t}+\frac{1}{4}\epsilon\Omega_c\frac{k_z^2}{\omega_{\mathbf{k},0}^2}e^{i[2\omega_{\mathbf{k},0}-\Omega_c]t}
    -\frac{h_+}{8i}\Omega_g\frac{k_x^2-k_y^2}{\omega_{\mathbf{k},0}^2} e^{i[2\omega_{\mathbf{k},0}+\Omega_c]t}-\frac{h_+}{8i}\Omega_g\frac{k_x^2-k_y^2}{\omega_{\mathbf{k},0}^2} e^{i[2\omega_{\mathbf{k},0}-\Omega_c]t}\nonumber\\
   \nonumber  &-\frac{\epsilon h_+}{8}\frac{k_z^2(k_x^2-k_y^2)}{\omega_{\mathbf{k},0}^3} \frac{\Omega_c^2+\Omega_g^2}{\Omega_c\Omega_g}e^{i[2\omega_{\mathbf{k},0}+(\Omega_c+\Omega_g)]t}
   -\frac{\epsilon h_+}{8}\frac{k_z^2(k_x^2-k_y^2)}{\omega_{\mathbf{k},0}^3} \frac{\Omega_c+\Omega_g}{\Omega_c\Omega_g}e^{i[2\omega_{\mathbf{k},0}-(\Omega_c-\Omega_g)]t}\nonumber\\
      &+\frac{\epsilon h_+}{8}\frac{k_z^2(k_x^2-k_y^2)}{\omega_{\mathbf{k},0}^3} \frac{\Omega_c+\Omega_g}{\Omega_c\Omega_g}e^{i[2\omega_{\mathbf{k},0}-(\Omega_g-\Omega_c)]t}
   +\frac{\epsilon h_+}{8}\frac{k_z^2(k_x^2-k_y^2)}{\omega_{\mathbf{k},0}^3} \frac{\Omega_c-\Omega_g}{\Omega_c\Omega_g}e^{i[2\omega_{\mathbf{k},0}-(\Omega_c+\Omega_g)]t}.
   \label{eq:Bkk}
\end{align}

By choosing specific resonance conditions and supposing that no other mode will couple with it, it is possible to see that only one term for each resonance condition sets its exponential amplitude to unity ($e^{i[2\omega_{\mathbf{k},0}\pm (\Omega_c\pm\Omega_g)]}\to 1$). We then allow to approximate Eq.~\eqref{eq:Bkk} to  
\begin{equation}
 B_{\mathbf{k},\mathbf{k}}(t)\to 
\begin{cases}
\frac{\epsilon}{4}\frac{\Omega_c k_z^2}{\omega_{\mathbf{k},0}^2},&\qquad\text{for}\qquad 2\omega_{\mathbf{k},0}=\Omega_c,\\
 \frac{ih_+}{8}\frac{\Omega_g(k_x^2-k_y^2)}{\omega_{\mathbf{k},0}^2} ,&\qquad\text{for}\qquad 2\omega_{\mathbf{k},0}=\Omega_g,\\
 \frac{\epsilon h_+}{8}\frac{k_z^2(k_x^2-k_y^2)}{\omega_{\mathbf{k},0}^3} \frac{\Omega_c^2+\Omega_g^2}{\Omega_c\Omega_g},&\qquad\text{for}\qquad 2\omega_{\mathbf{k},0}=|\Omega_c\pm\Omega_g|,
\end{cases}  
\end{equation}
which is justified in light of the fact that all the other terms with highly oscillating exponential amplitudes average to zero after a time average procedure.

We then have three different resonance conditions:
\begin{subequations}
\label{eq:H3}
\begin{align}
  \omega_{\mathbf{k}}=\frac{\Omega_c}{2}:&\hspace{0.3cm} \hat{H}=-\frac{ g_0}{2}\left(\hat{a}_{\mathbf{k}}^{\dagger 2}+\hat{a}_{\mathbf{k}}^2\right)\hspace{0.3cm} \text{with}\hspace{0.3cm} g_0=\left|\frac{\epsilon}{4}\frac{\Omega_c k_z^2}{\omega_{\mathbf{k},0}^2}\right|,\\
  \omega_{\mathbf{k}}=\frac{\Omega_g}{2}:&\hspace{0.3cm}  \hat{H}=-\frac{ g_1}{2}\left(\hat{a}_{\mathbf{k}}^{\dagger 2}+\hat{a}_{\mathbf{k}}^2\right)\hspace{0.3cm} \text{with}\hspace{0.3cm} g_1=\left|\frac{h_+}{8}\frac{\Omega_g}{\omega_{\mathbf{k},0}^2}(k_x^2-k_y^2)\right|,\\
  \omega_{\mathbf{k}}=\frac{|\Omega_c\pm\Omega_g|}{2}:&\hspace{0.3cm}  \hat{H}=\frac{ i g_2}{2}\left(\hat{a}_{\mathbf{k}}^{\dagger 2}-\hat{a}_{\mathbf{k}}^2\right)\hspace{0.3cm} \text{with}\hspace{0.3cm} g_2=\left|\frac{\epsilon h_+}{8}\frac{(k_x^2-k_y^2)k_z^2}{\omega_{\mathbf{k},0}^3} \frac{\Omega_c^2+\Omega_g^2}{\Omega_c\Omega_g}\right|.
\end{align}
\end{subequations}
\subsubsection{For the $B_{\mathbf{k},\mathbf{j}}(t)$ terms}

Again, by using Eqs.~\eqref{eq:wk}, one can prove the following expansions,
\begin{align}
    G_{\mathbf{k}\mathbf{j}}(t)&=(-1)^{j_z-k_z}\frac{2k_z j_z}{j_z^2-k_z^2}\epsilon \Omega_c\cos(\Omega_c t)\delta_{k_xj_x}\delta_{k_y,j_y}\nonumber\quad \text{for} \quad j_z \neq k_z,\\
    \sqrt{\frac{\omega_{\mathbf{k}}(t)}{{\omega_{\mathbf{j}}(t)}}}&=\sqrt{\frac{\omega_{\mathbf{k},0}}{\omega_{\mathbf{j},0}}}\Bigg[1+\frac{h_+}{4}\left(\frac{k_x^2-k_y^2}{\omega_{\mathbf{k},0}^2}-\frac{j_x^2-j_y^2}{\omega_{\mathbf{j},0}^2}\right)\cos(\Omega_g t)\Bigg]+\mathcal{O}(\epsilon,h^2)
\end{align}

In this situation, one has the following
\begin{align}
\label{eq:eOkOj}
    \mu_{\mathbf{k}\mathbf{j}}(t)=(-1)^{j_z-k_z}\frac{2k_z j_z}{j_z^2-k_z^2}\sqrt{\frac{\omega_{\mathbf{k},0}}{\omega_{\mathbf{j},0}}}\Omega_c\epsilon\cos(\Omega_ct)\Bigg[1+\frac{h_+}{4}\left(\frac{k_z^2-j_z^2}{\omega_{\mathbf{k},0}^2\omega_{\mathbf{j},0}^2}\right)(k_x^2-k_y^2)\cos(\Omega_g t)\Bigg]\delta_{k_xj_x}\delta_{k_y,j_y},
\end{align}
where we have used the identity $\omega_{\mathbf{k},0}^2-\omega_{\mathbf{j},0}^2=k_z^2-j_z^2$ whenever $k_x=j_x$ and $k_y=j_y$. As a result, one has
\begin{align}
    \frac{1}{2}\left[\mu_{\mathbf{k}\mathbf{j}}+\mu_{\mathbf{j}\mathbf{k}}\right]&=(-1)^{j_z-k_z}\Omega_c\epsilon\cos(\Omega_ct)\frac{k_z j_z}{j_z^2-k_z^2}\frac{\omega_{\mathbf{k},0}-\omega_{\mathbf{j},0}}{\sqrt{\omega_{\mathbf{k},0}\omega_{\mathbf{j},0}}}\delta_{k_xj_x}\delta_{k_y,j_y}\nonumber\\
    \label{eq:mukjjk}&+(-1)^{j_z-k_z}\frac{\epsilon h_+}{4}\Omega_c\frac{k_z j_z}{\omega_{\mathbf{k},0}^2\omega_{\mathbf{j},0}^2}\frac{\omega_{\mathbf{k},0}+\omega_{\mathbf{j},0}}{\sqrt{\omega_{\mathbf{k},0}\omega_{\mathbf{j},0}}}(k_x^2-k_y^2)\cos(\Omega_ct)\cos(\Omega_g t)\delta_{k_xj_x}\delta_{k_y,j_y}.
\end{align}
Using the fusion expressions~\eqref{eq:eOkOj} and~\eqref{eq:mukjjk} one obtains 
\begin{align*}
    B_{\mathbf{k}\mathbf{j}}&={\frac{1}{2}\left[\mu_{\mathbf{k}\mathbf{j}}+\mu_{\mathbf{j}\mathbf{k}}\right]}{e^{i\left[\Theta_{\mathbf{k}}+\Theta_{\mathbf{j}}\right]}}\approx(-1)^{j_z-k_z}\Omega_c\epsilon\Bigg\{\frac{k_z j_z}{j_z^2-k_z^2}\frac{\omega_{\mathbf{k},0}-\omega_{\mathbf{j},0}}{\sqrt{\omega_{\mathbf{k},0}\omega_{\mathbf{j},0}}}\cos(\Omega_ct)e^{i\left[\omega_{\mathbf{k},0}+\omega_{\mathbf{j},0}\right]t}\\
    &+\frac{h_+}{4}\frac{k_z j_z}{\omega_{\mathbf{k},0}^2\omega_{\mathbf{j},0}^2}\frac{\omega_{\mathbf{k},0}+\omega_{\mathbf{j},0}}{\sqrt{\omega_{\mathbf{k},0}\omega_{\mathbf{j},0}}}(k_x^2-k_y^2)\cos(\Omega_ct)\cos(\Omega_g t)e^{i\left[\omega_{\mathbf{k},0}+\omega_{\mathbf{j},0}\right]t}\\
    &-\frac{ih_+}{2\Omega_g}\frac{k_z j_z}{j_z^2-k_z^2}\frac{\omega_{\mathbf{k},0}-\omega_{\mathbf{j},0}}{\sqrt{\omega_{\mathbf{k},0}\omega_{\mathbf{j},0}}}\frac{\omega_{\mathbf{k},0}+\omega_{\mathbf{j},0}}{\omega_{\mathbf{k},0}\omega_{\mathbf{j},0}}(k_x^2-k_y^2)\cos(\Omega_ct)\sin(\Omega_g t)e^{i\left[\omega_{\mathbf{k},0}+\omega_{\mathbf{j},0}\right]t}\Bigg\}\delta_{k_xj_x}\delta_{k_y,j_y}.
    \end{align*}
    
Then we have
\begin{align}
   \nonumber B_{\mathbf{k}\mathbf{j}}&\approx \frac{(-1)^{j_z-k_z}}{2}\Omega_c\epsilon\frac{k_z j_z}{j_z^2-k_z^2}\frac{\omega_{\mathbf{k},0}-\omega_{\mathbf{j},0}}{\sqrt{\omega_{\mathbf{k},0}\omega_{\mathbf{j},0}}}e^{i[\omega_{\mathbf{k},\mathbf{j}}+\Omega_c]t}+\frac{(-1)^{j_z-k_z}}{2}\Omega_c\epsilon\frac{k_z j_z}{j_z^2-k_z^2}\frac{\omega_{\mathbf{k},0}-\omega_{\mathbf{j},0}}{\sqrt{\omega_{\mathbf{k},0}\omega_{\mathbf{j},0}}}e^{i[\omega_{\mathbf{k},\mathbf{j}}-\Omega_c]t}\\
   \nonumber&-(-1)^{j_z-k_z}\frac{h_+\epsilon}{8\omega_{\mathbf{k},0}\omega_{\mathbf{j},0}}\frac{\Omega_c}{\Omega_g}\frac{(k_x^2-k_y^2)k_z j_z}{\sqrt{\omega_{\mathbf{k},0}\omega_{\mathbf{j},0}}}\left(1+\frac{\Omega_g}{2}\frac{\omega_{\mathbf{k},0}+\omega_{\mathbf{j},0}}{\omega_{\mathbf{k},0}\omega_{\mathbf{j},0}}\right)e^{i[\omega_{\mathbf{k},\mathbf{j}}+(\Omega_c+\Omega_g)]t}\\
  \nonumber &+(-1)^{j_z-k_z}\frac{h_+\epsilon}{8\omega_{\mathbf{k},0}\omega_{\mathbf{j},0}}\frac{\Omega_c}{\Omega_g}\frac{(k_x^2-k_y^2)k_z j_z}{\sqrt{\omega_{\mathbf{k},0}\omega_{\mathbf{j},0}}}\left(1-\frac{\Omega_g}{2}\frac{\omega_{\mathbf{k},0}+\omega_{\mathbf{j},0}}{\omega_{\mathbf{k},0}\omega_{\mathbf{j},0}}\right)e^{i[\omega_{\mathbf{k},\mathbf{j}}-(\Omega_c-\Omega_g)]t}\\
  \nonumber &+(-1)^{j_z-k_z}\frac{h_+\epsilon}{8\omega_{\mathbf{k},0}\omega_{\mathbf{j},0}}\frac{\Omega_c}{\Omega_g}\frac{(k_x^2-k_y^2)k_z j_z}{\sqrt{\omega_{\mathbf{k},0}\omega_{\mathbf{j},0}}}\left(1+\frac{\Omega_g}{2}\frac{\omega_{\mathbf{k},0}+\omega_{\mathbf{j},0}}{\omega_{\mathbf{k},0}\omega_{\mathbf{j},0}}\right)e^{i[\omega_{\mathbf{k},\mathbf{j}}-(\Omega_g-\Omega_c)]t}\\
   \label{eq:Bkj}&+(-1)^{j_z-k_z}\frac{h_+\epsilon}{8\omega_{\mathbf{k},0}\omega_{\mathbf{j},0}}\frac{\Omega_c}{\Omega_g}\frac{(k_x^2-k_y^2)k_z j_z}{\sqrt{\omega_{\mathbf{k},0}\omega_{\mathbf{j},0}}}\left(1+\frac{\Omega_g}{2}\frac{\omega_{\mathbf{k},0}+\omega_{\mathbf{j},0}}{\omega_{\mathbf{k},0}\omega_{\mathbf{j},0}}\right)e^{i[\omega_{\mathbf{k},\mathbf{j}}-(\Omega_c+\Omega_g)]t}
    \end{align}
    
Again, by choosing specific resonance conditions and supposing that no other mode will couple with it, we can approximate Eq.~\eqref{eq:Bkj} as
\begin{equation}
 B_{\mathbf{k},\mathbf{k}}(t)\to 
\begin{cases}
 (-1)^{j_z-k_z}\frac{h_+\epsilon}{8\omega_{\mathbf{k},0}\omega_{\mathbf{j},0}}\frac{\Omega_c}{\Omega_g}\frac{(k_x^2-k_y^2)k_z j_z}{\sqrt{\omega_{\mathbf{k},0}\omega_{\mathbf{j},0}}}\left(1+\frac{\Omega_g}{2}\frac{\Omega_c\pm\Omega_g}{\omega_{\mathbf{k},0}\omega_{\mathbf{j},0}}\right),&\hspace{0.3cm}\text{for}\hspace{0.3cm} \omega_{\mathbf{k},0}+\omega_{\mathbf{j},0}=\Omega_g\pm\Omega_c,\\
(-1)^{j_z-k_z}\frac{h_+\epsilon}{8\omega_{\mathbf{k},0}\omega_{\mathbf{j},0}}\frac{\Omega_c}{\Omega_g}\frac{(k_x^2-k_y^2)k_z j_z}{\sqrt{\omega_{\mathbf{k},0}\omega_{\mathbf{j},0}}}\left(1-\frac{\Omega_g}{2}\frac{\Omega_c-\Omega_g}{\omega_{\mathbf{k},0}\omega_{\mathbf{j},0}}\right),&\hspace{0.3cm}\text{for}\hspace{0.3cm} \omega_{\mathbf{k},0}+\omega_{\mathbf{j},0}=\Omega_c-\Omega_g,
\end{cases}  
\end{equation}

We then have two more resonance conditions
\begin{eqnarray}
\label{eq:H2}
\omega_{\mathbf{k},0}+\omega_{\mathbf{j},0}&=&\Omega_g\pm\Omega_c: \hspace{0.3cm} \hat{H}=\frac{ i g_1}{2}\left(\hat{a}_{\mathbf{k}}^{\dagger}\hat{a}_{\mathbf{j}}^{\dagger}-\hat{a}_{\mathbf{k}}\hat{a}_{\mathbf{j}}\right), \nonumber\\ 
&\text{with}&\hspace{0.3cm} g_1=\left|\frac{h_+\epsilon}{8\omega_{\mathbf{k},0}\omega_{\mathbf{j},0}}\frac{\Omega_c}{\Omega_g}\frac{k_z j_z(k_x^2-k_y^2)}{\sqrt{\omega_{\mathbf{k},0}\omega_{\mathbf{j},0}}}\left(1+\frac{\Omega_g}{2}\frac{\Omega_c\pm\Omega_g}{\omega_{\mathbf{k},0}\omega_{\mathbf{j},0}}\right)\right|,\nonumber \\
\omega_{\mathbf{k},0}+\omega_{\mathbf{j},0}&=&\Omega_c-\Omega_g: \  \hat{H}=\frac{ i g_2}{2}\left(\hat{a}_{\mathbf{k}}^{\dagger}\hat{a}_{\mathbf{j}}^{\dagger}-\hat{a}_{\mathbf{k}}\hat{a}_{\mathbf{j}}\right),\nonumber \\ 
&\text{with}&\hspace{0.3cm} g_2=\left|\frac{ih_+\epsilon}{8\omega_{\mathbf{k},0}\omega_{\mathbf{j},0}}\frac{\Omega_c}{\Omega_g}\frac{k_z j_z(k_x^2-k_y^2)}{\sqrt{\omega_{\mathbf{k},0}\omega_{\mathbf{j},0}}}\left(1-\frac{\Omega_g}{2}\frac{\Omega_c-\Omega_g}{\omega_{\mathbf{k},0}\omega_{\mathbf{j},0}}\right)\right|.
\end{eqnarray}

Using the Hamiltonian operators Eqs.~\eqref{eq:H3} and~\eqref{eq:H2} together with formula~\eqref{eq:Np} we can finally calculate the number of particles created in the vacuum in each resonance condition:
\begin{subequations}
\begin{align}
  \omega_{\mathbf{k}}=\frac{\Omega_g}{2}:&\qquad  N_{\mathbf{k}}(T)=\sinh^2\left|\frac{h_+}{8}\frac{\Omega_g}{\omega_{\mathbf{k},0}^2}(k_x^2-k_y^2)\right|,\label{eq:app:NkT1}\\
  \omega_{\mathbf{k}}=\frac{|\Omega_c\pm\Omega_g|}{2}:&\qquad N_{\mathbf{k}}(T)=\sinh^2\left|\frac{\epsilon h_+}{8}\frac{(k_x^2-k_y^2)k_z^2}{\omega_{\mathbf{k},0}^3} \frac{\Omega_c^2+\Omega_g^2}{\Omega_c\Omega_g}\right|,\label{eq:app:NkT2}\\
\omega_{\mathbf{k}}+\omega_{\mathbf{j}}=\Omega_g\pm\Omega_c:&\qquad N_{\mathbf{k}}(T)=\sinh^2\left|\frac{h_+\epsilon}{16\omega_{\mathbf{k},0}\omega_{\mathbf{j},0}}\frac{\Omega_c}{\Omega_g}\frac{(k_x^2-k_y^2)k_z j_z}{\sqrt{\omega_{\mathbf{k},0}\omega_{\mathbf{j},0}}}\left(1+\frac{\Omega_g}{2}\frac{\Omega_c\pm\Omega_g}{\omega_{\mathbf{k},0}\omega_{\mathbf{j},0}}\right)\right|,\label{eq:app:NkT3}\\
\omega_{\mathbf{k}}+\omega_{\mathbf{j}}=\Omega_c-\Omega_g:&\qquad N_{\mathbf{k}}(T)=\sinh^2\left|\frac{h_+\epsilon}{16\omega_{\mathbf{k},0}\omega_{\mathbf{j},0}}\frac{\Omega_c}{\Omega_g}\frac{(k_x^2-k_y^2)k_z j_z}{\sqrt{\omega_{\mathbf{k},0}\omega_{\mathbf{j},0}}}\left(1-\frac{\Omega_g}{2}\frac{\Omega_c-\Omega_g}{\omega_{\mathbf{k},0}\omega_{\mathbf{j},0}}\right)\right|,\label{eq:app:NkT4}
\end{align}
\end{subequations}
whenever $k_x=j_x$ and $k_y=j_y$.


\end{document}